%% file: KrauseAlexander.tex
\documentclass[useAMS,usenatbib,usegraphicx]{mn2e}

\include{def}
\title[Multi-phase turbulence in jet cocoons]
{Simulations of multi-phase turbulence in jet cocoons}
\author[M. Krause and P. Alexander]{M. Krause\thanks{E-mail:
M.Krause@mrao.cam.ac.uk} and P. Alexander\\
Astrophysics Group, Cavendish Laboratory, Cambridge CB3 0HE, UK}
\begin{document}

\date{Accepted \date. Received 2005 August 10}

\pagerange{\pageref{firstpage}--\pageref{lastpage}} \pubyear{2006}

\maketitle

\label{firstpage}

\begin{abstract}
The interaction of optically emitting clouds with warm X-ray gas and hot, tenuous 
radio plasma
in radio jet cocoons is modelled by 2D compressible hydrodynamic simulations. 
The initial setup is the Kelvin-Helmholtz instability at a contact surface of 
density contrast $10^4$. The denser medium contains clouds of higher density. 
Optically thin radiation is realised via a cooling source term.  
The cool phase effectively extracts energy from the other gas which is both, 
radiated away and used for acceleration of the cold phase. 
This increases the system's cooling rate substantially and leads to a 
massively amplified cold mass dropout. We show that it is feasible, given small 
seed clouds of order 100~$M_\mathrm{\sun}$, that all of the optically emitting gas 
in a radio jet cocoon may be produced by this mechanism on the propagation 
timescale of the jet. The mass is generally distributed as $T^{-1/2}$ with 
temperature, with a prominent peak at 14,000~K. This peak is likely to be 
related to the counteracting effects of shock heating and a strong rise in the 
cooling function. The volume filling factor of cold gas in this peak is of the 
order $10^{-5}$ to $10^{-3}$ and generally increases during the simulation time.\\ 
The simulations tend towards an isotropic scale free Kolmogorov-type energy spectrum 
over the simulation
timescale. We find the same Mach-number density relation
as \citet{KN04} and show that this relation may explain the velocity widths 
of emission lines associated with high redshift radio galaxies, if the environmental
temperature is lower, or the jet-ambient density ratio is less extreme than in their
low redshift counterparts.
\end{abstract}

\begin{keywords}
hydrodynamics -- instabilities -- turbulence -- galaxies: jets --
methods: numerical.
\end{keywords}

\section{Introduction}
Extragalactic radio jets have been observed to interact strongly with their 
environment. This is evident not only from their often fat radio cocoons
but also from the cospatial optical and X-ray emission (detailed below). 
This is observational evidence for what is expected from theory, namely
the co-existence of hot magnetised cocoon plasma, warm X-ray gas, and cold,
possibly star forming clouds, all interacting with one another.

This article is particularly 
stimulated by observations of radio galaxies at a redshift 
of about one
(e.g. \citealt{HippMeis92}; \citealt{MeisHipp92}; \citealt*{BLR96,BLR97,Neesea97};
\citealt{Bestea98}; \citealt*{BLR98}; \citealt{Bestea99}; \citealt*{Neesea99};
\citealt{Bestea00a}; \citealt*{BRL00a,BRL00b,Best00}; \citealt{Inskea02b,Inskea02c,Inskea03,Inskea05}).
Clear trends are observed: the smaller radio 
sources have the larger emission line
regions, the greater line widths, and are more consistent with shock ionisation.
This is further evidence for the tight connection between the 
gas phases present.

The traditional viewpoint has been to study the individual phases
separately, analogously to interstellar medium (ISM) 
turbulence theory \citep{ES04}.
As for the case of the ISM, new insights can be expected from direct 
hydrodynamical simulation of the interaction.
These interactions are so far unknown. The open questions
include energy, momentum, and mass transfer between the phases.
The origin of the cold phase is unclear \citep[apart from the merger cases, compare][]{Neesea99}: 
a common idea is that pre-existing 
clouds are activated for example by the bow shock or photo-ionised by a central 
quasar. A possible problem of this model might be the low visibility of clouds far from
the radio emission, which would still be expected to be photo-ionised
 \citep{Inskea02c}. It is rather unlikely that much of the cool 
gas is drawn from the host galaxy, because the solid angle occupied by the 
beam is quite small, and except at the very end of the jet the densest clouds 
are likely to pass through the contact surface into the cocoon, where the flow
if anything moves the clouds back towards the galaxy 
\citep[e.g.][]{AP96}. Given the probably similar masses of warm and cold phase
(see Section~\ref{el}),
it appears possible that the latter cooled and condensed from the former.
This seems difficult for sources in nearby galaxy clusters, but may work 
at higher redshift, where the cluster gas is expected to be cooler, 
and therefore
to have a shorter cooling time. Weak shocks caused by the jet may further
lower the cooling time. This idea would explain the dependence of the 
appearance of cold gas on redshift (see Section~\ref{el}).

A better understanding of these inter-relations would not 
only advance 
radio source physics, but bear the potential of using jets and the associated 
emission as probes of the surrounding (cluster) gas. 
As cluster gas properties at 
redshifts of one and beyond are still hard to determine with present day X-ray 
telescopes \cite[e.g.][]{Andrea05}, this may give important constraints, with 
implications for cosmology and galaxy formation.

We plan to address the problem in three ways. 
Here we concentrate on simulations that contain the three phases
from the beginning and study their interaction in kpc-sized box simulations.
Clues from large scale 
jet propagation and investigation of the possibility of condensing the 
cold phase entirely out of the warm one are deferred to future publications.

The three phases are reviewed individually in Section~\ref{3phases}.
Section \ref{turbu} discusses the relation to turbulence research, the 
simulations are presented in Section~\ref{sims} and discussed in 
Section~\ref{disc}.

\section{The three phases present in radio galaxies}\label{3phases}
\subsection{Hot magnetised radio plasma}\label{radio}
Many of the more powerful FR~II sources \citep{FR74,Gilbea04} possess 
fat radio cocoons. Provided the jets are light compared to the ambient gas,
beam plasma that has passed the terminal shock region
assembles in roughly cylindrical cocoons \citep[e.g.][]{Normea82}.
According to momentum balance,
the head advance speed $v_\mathrm{h}$
drops with the beam density $\rho_\mathrm{j}$ as:
\begin{equation}
v_\mathrm{h}=\sqrt{\eta\epsilon} v_\mathrm{j}\, ,
\end{equation} 
where $\eta=\rho_\mathrm{j}/\rho_0$, $\rho_0$ is the ambient density,
$v_\mathrm{j}$ is the jet speed, and $\epsilon$
the model dependent thrust distribution 
factor ($<1$). Therefore, the lighter the jets, the slower the head advance speed,
and the wider the cocoons have to be in order to store the shocked beam plasma.
Simulations still struggle to reproduce the observed cocoon widths, which 
at least in some cases require $\eta \la 10^{-4}$ \citep{Krause2005a},
in agreement with analytical considerations \citep[e.g.][]{AP96}.
The jet velocity must be a reasonable fraction of the speed of light -- jet 
prominence \citep{Hardea99} and arm-length asymmetries \citep{AL04}
suggest $v_\mathrm{j} \ga c/2$ whereas a bulk Lorentz factor of about 10 
is suggested by X-ray--Compton studies \citep[][and references therein]{Tavea04}
and the morphology and size of the bow shock in Cygnus~A \citep{Krause2005a}.
These conditions in the jet imply, after the jet shock, a very hot tenuous
cocoon plasma with an effective temperature of order
\begin{equation}
T_\mathrm{c}= 3 \times 10^{11} \left(\frac{v_\mathrm{j}}{c/2}\right)^2 \, \mathrm{K} \,\, .
\end{equation}

\begin{figure}
\centering
\includegraphics[height=0.45\textwidth, angle=-90]{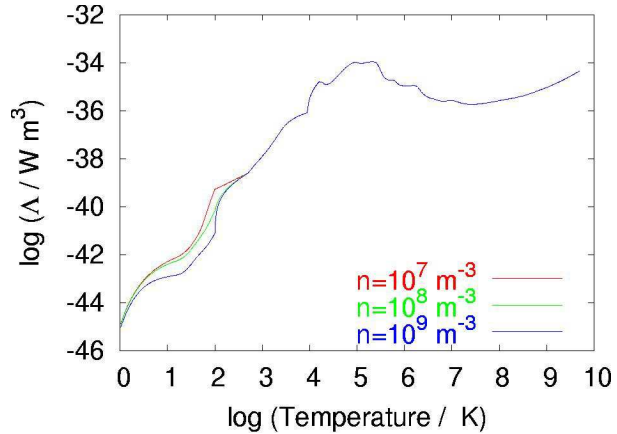}
\caption{Employed cooling curve. The low temperature part is density dependent.}
\label{coolcurve}
\end{figure}
\subsection{Warm X-ray gas}\label{xray}
The radio cocoons are in rough pressure equilibrium with the surrounding 
shocked X-ray gas at, typically, $10^7$--$10^8$~K. 
Therefore, and for dynamical reasons (compare above) the typical density 
ratio between cocoon plasma and surrounding X-ray gas should be about 
$10^{-4}$.
The dividing contact surface is subject to Rayleigh-Taylor and 
Kelvin-Helmholtz instabilities, although it may be stabilised 
by deceleration \citep{Alex02,Krause2005a} and magnetic fields 
\citep{mypap01a} for some time. Hydrodynamic simulation indicates 
that a few percent of the gas mass pushed aside by the radio cocoon 
is entrained back into the 
cocoon region by these instabilities, typically of
the order $10^9 M_{\rm \mathrm{\sun}}$ for sources of size 100~kpc \citep{Krause2005a}.
This picture is confirmed by radio 
observations \citep[e.g.][]{Gilbea04}. While the regions close 
to the hotspots of FR~II sources are usually sharply bound, 
indicating a stable contact surface, towards the centres the 
emission diffuses away and gets filamentary, 
indicating entrainment of the surrounding shocked 
ambient gas. The latter is very obvious from the high dynamic range 
observations of Cygnus~A of \citet{PDC84}. 
The entrained X-ray gas has been resolved by Chandra observations of 
the same source \citep{Sea01}. In the following, this will be called the
warm phase in contrast to the hot, radio-emitting cocoon-plasma.

\subsection{Cold emission line gas}\label{el}
A third gas component is revealed by the alignment effect
\citep[for a review see][]{MC93}. Beyond a redshift of $0.6$,
 optical emission is aligned with the radio jet and often cospatial 
with the cocoon. This emission is indicative 
of comparatively cold ($T\approx 10^4$~K) gas. 
The emission is partly polarised to varying degree, which 
is interpreted as a scattering contribution of light from a hidden 
quasar by dust grains \citep[e.g.][and references therein]{SolInea04}.
Additionally, nebular emission \citep{Inskea02b,Inskea02c}
and newly formed stars \citep[e.g.][]{BLR96,DEA97} may contribute to the 
continuum light \citep[but see also][]{Neesea99}. The emission line luminosity
is well correlated with the radio power.
Such a description sounds much like the ordinary 
inter-stellar medium (ISM) 
in galaxies. The ISM is traditionally regarded as a multi-phase medium
itself. However, recent advances in turbulence research have revealed rapid 
transformation between clouds and inter-cloud medium, wherefore
\citet{ES04} drop the distinction. We adopt this viewpoint and regard all the
gas that is able to cool significantly by emission on a reasonable time-scale
(see below) as the third phase, which will be referred to as the cold phase.

\citet{BMcC00} give a range of $10^8$--$10^{10} M_{\rm \mathrm{\sun}}$
for the cold gas component for radio galaxies at $z\approx 1$.
This is likely to be much more than the radio plasma by a factor of at least 
a hundred, given the density ratio discussed above, and about the same
as the entrained X-ray gas, as inferred from simulation.

\section{Connection to turbulence theory}\label{turbu} 
'The appearance of instability in viscous shear flows usually results
in the nonlinear outcome: the onset of turbulence' \cite[][p 119]{Shu92ii}.
FR~I jets have been recognised as turbulent flows early on 
\cite[e.g.][]{Bick84,Kom90a}. The cocoons of FR~II jets are even more likely 
to be turbulent, since these shear flows probably have low velocities
with respect to the surrounding shocked ambient gas. This has already been
mentioned in the early simulation papers \citep[e.g.][]{KM88}. The 
appearance of smaller and smaller vortices with increasing resolution
has actually been observed in a resolution study by \citet{mypap01a},
who varied the resolution of their simulations by up to a factor of 40.
Hence, the turbulent nature of the cocoons of FR~II sources is evident, and the
recent literature treats it as a matter of course \citep{SBS02,CO02b}.

An excellent review of interstellar turbulence can be found in \citet{ES04}.
They point out that compressible turbulence, which is relevant here,
is still not fully understood. For example, an inertial range where kinetic
energy is transferred in a scale-free manner (the Kolmogorov law) to
smaller scales (or larger scales as may be the case for 2D turbulence) 
is not expected
a priori. The character of turbulence is furthermore dependent on 
magnetic fields and self-gravity. 
Since turbulence is a phenomenon which covers a wide range of physical scales,
resolution in numerical simulations is an issue. In this paper we present simulations
which while, by necessity, are of limited resolution, we believe offer 
improved physical insight.
Some aspects of the problem we address have been studied by other authors
\citep[e.g.][]{KN02,KN04,MKR02}, thereby allowing some consistency check 
on our results. 

Turbulence quickly decays, if not driven constantly. Its properties,
however, may depend on the driving mechanism \citep{ES04}. 
Many turbulence simulations use large scale driving by source terms.
In the simulations presented below, the energy is initially stored mainly in 
kinetic energy, and transferred slowly to turbulent motions
by the Kelvin-Helmholtz instability. Because the linear
growth times of these instabilities depend on the length 
scale (Equations~(\ref{khtime}),
the driving starts on small scales and involves larger scales the longer the 
simulation is run for. This is designed to reproduce the conditions in jet cocoons as 
closely as possibly, but it is different to most simulations in the literature. 

There are a number of general results concerning compressible turbulence
from theory and simulation which we expect to find reflected in our simulations.
Turbulence theory finds log-normal density probability distribution 
functions (PDF) for isothermal turbulence \citep{ES04}: power law tails are 
expected for specific heat ratios other than one. The volume filling factor 
as a function of density is determined to be a power law from simulation. 
The velocity PDF is expected to be Gaussian from analytic work but is found 
to be more exponential from simulations. In the cold phase, the average
Mach number grows with density as $\bar{M}\, \propto \rho^{1/2}$ \citep{KN04}.

\section{The simulations}\label{sims}
\begin{figure}
\centering
\includegraphics[width=0.45\textwidth]{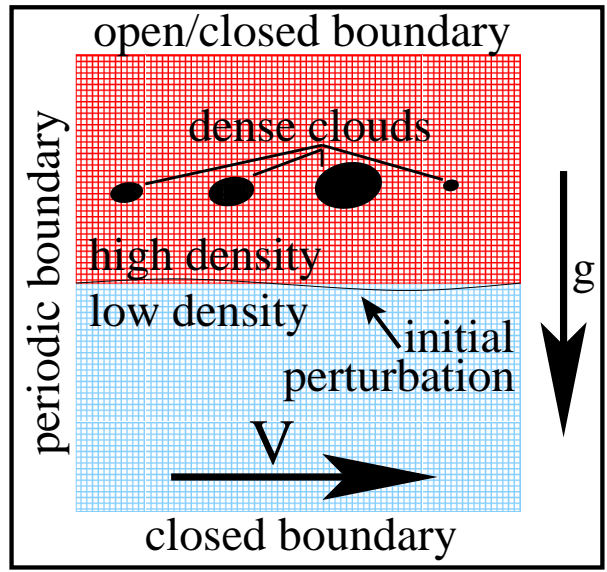}
\caption{Sketch of the setup for all simulations. The three phases soon mix 
due to fluid instabilities, and form a turbulent region powered by the initial
gravitational and kinetic energy.}
\label{simsketch}
\end{figure}
Hydrodynamic simulations were performed in order to study the turbulent 
interaction of the three phases. The equations for mass ($\rho$~d$x^3$), 
momentum ($\rho {\bf v}$~d$x^3$), and internal energy ($e$~d$x^3$) are
\begin{eqnarray}
\dt{\rho}+\nabla \cdot (\rho {\bf v}) &=& 0 \\
\dt{\rho {\bf v}} + \nabla (\rho {\bf v v}) &=& -\nabla p -g \rho\\
\dt{e} + \nabla (e {\bf v}) &
	=& -p \nabla \cdot {\bf v} - \frac{\rho^2}{m_\mathrm{p}^2} \Lambda,
\end{eqnarray}
where $m_\mathrm{p}$ is the proton mass and the system is closed by the equation
of state $p=(\gamma-1) e$ with $\gamma=5/3$ being the ratio of specific heats.
These equations were evolved in two Cartesian dimensions with the code NIRVANA 
\citep{ZY97}. This code is second order accurate in space and time, and is able 
to resolve shocks via artificial viscosity (compressible hydrodynamics).

Cooling is treated as a source term in the energy equation. The cooling
curve was adopted from \citet{basson02}. For temperatures
between $10^4$K and $10^{8.5}$K, the \citet{SD93} cooling curve
for solar metallicity was used.
Above that temperature the cooling is due to bremsstrahlung:
$\Lambda=2.05\times 10^{-40}\sqrt{T}  (1+4.4 \times10^{-10} T)$~Jm$^3$/s, where the second
term in the bracket is the relativistic correction. For one run, the cooling curve
was cut below $10^4$~K, to emulate effects of photoionisation. For the others,
cooling by collisional de-excitation of molecular hydrogen 
\citep[][$10^2$K--$10^4$K]{Tegea97} and by emission lines from H$_2$, HD,
and CO molecules \citep[][$< 100$K]{Puyea99} were included.
This cooling curve is shown in Figure~\ref{coolcurve}.
Another source term is the constant gravity of $g=10^{-10} \mathrm{m\,s}^{-2}$.
The gravity does not influence the simulations by much but was included 
to make the simulations as realistic as possible.

We have also used the third order accurate Flash code, which is forced to 
conserve energy and momentum to check the results. We encounter numerical problems
with both codes, when regions of hugely differing densities approach each other.
This restricts the density ratios in the simulations.
It turned out that, with given computational resources, we could perform better
simulations with Nirvana. We could not reach enough resolution with Flash to 
produce reliable results in the parameter range of the presented Nirvana simulations.
Therefore, we discuss mainly our Nirvana simulations.

\begin{figure*}
\centering
\includegraphics[width=0.95\textwidth]{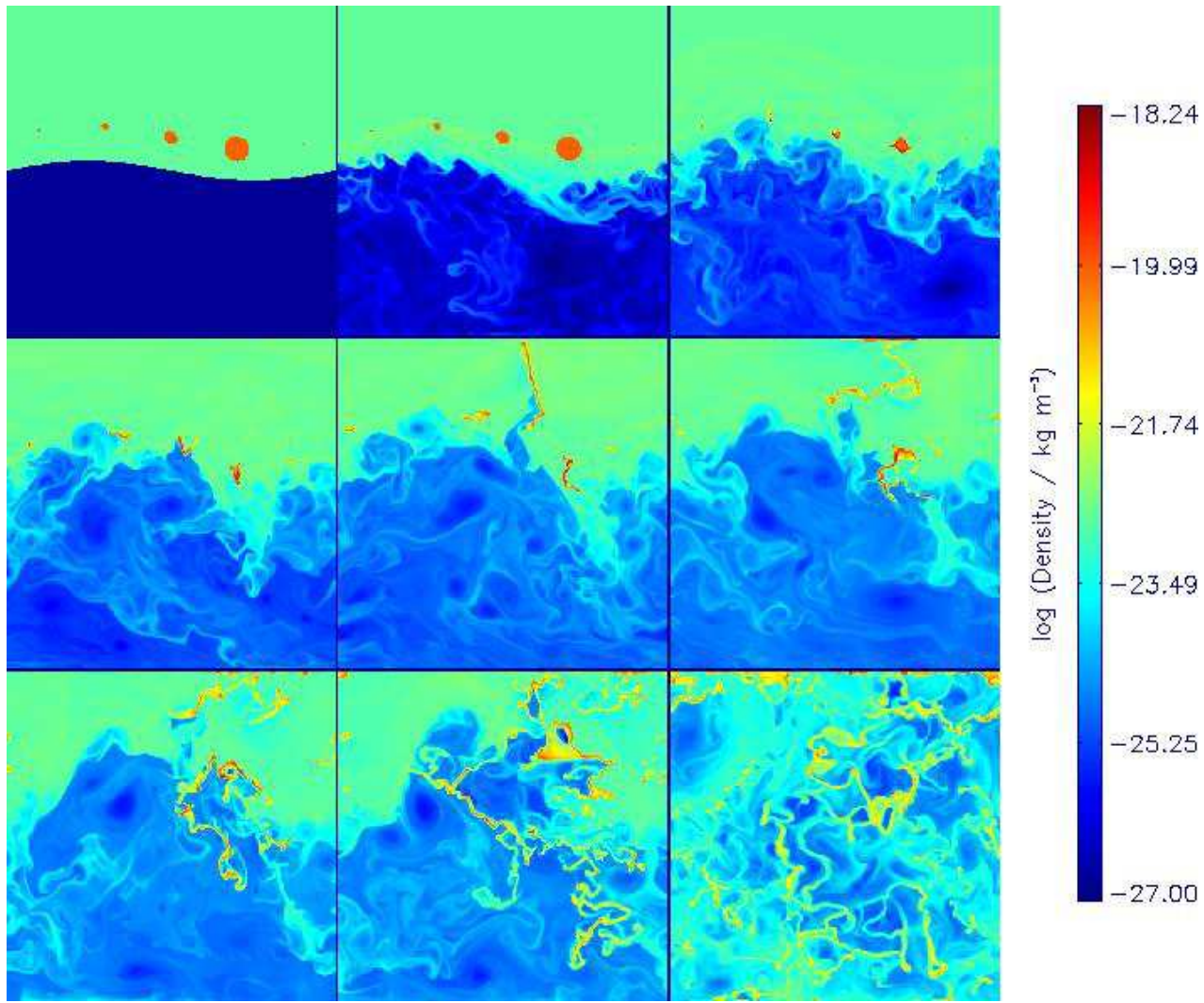}
\caption{Logarithmic density plots at nine representative simulation times
(0, 0.2, 0.9, 1.7, 2.4, 3.2, 4.1, 4.8, and 7.7 Myr) for RUN~7c4.}
\label{lgd7c4}
\end{figure*}
\subsection{Setup}
The initial setup of the simulations is shown in Figure~\ref{simsketch}.
The idea is to model the situation at the contact surface between jet cocoon
and shocked ambient gas. The upper half of the computational square box 
of size $[1\,\mathrm{kpc}\times 1\,\mathrm{kpc}]$, resolved by 
$[512\times512]$ cells, is initially filled with 
gas of density $10^{4} m_\mathrm{p}\,$m$^{-3}$. 
The pressure was set to decrease linearly with height, in order 
to achieve hydrostatic equilibrium. With a temperature of $2\times 10^6$K,
the cooling time is about 10~Myr, which is similar to our simulation time.
Five elliptical clouds with a very high density (see Table~\ref{simpars})  
and various radii (0.008,0.02,0.04,0.07,0.004 kpc) have been placed in the high
density region at a height of 0.6~kpc. This results in a total cold mass of 
$10^4$--$10^7 M_{\rm \mathrm{\sun}}$\,kpc$^{-3}$. 
For determination of this density and 
similar ones to follow, the cells 
are assumed to represent cubes of three equal sides. 
Hence, the values are upper limits due to the unknown 3D structure.
Observed emission line gas masses 
(see above) and radio cocoon sizes of about 
$V_\mathrm{coc}=\pi r_\mathrm{coc}^2 h \approx \pi (15\;\mathrm{kpc})^2 100\; 
\mathrm{kpc} \approx 3 \times 10^5 \mathrm{kpc}^3$ lead to cold gas masses
of $10^3$--$10^5 M_{\rm \mathrm{\sun}}$\,kpc$^{-3}$. 
We also include a control run without any clouds.
 
The lower half of the box contained low density gas 
(m$_\mathrm{p}\,$m$^{-3}$) at about the same pressure, which was adjusted 
experimentally for best numerical stability of the initial configuration.
This gas was given an initial velocity of $v=2\times10^7$~m$\,$s$^{-1}$, corresponding
to a Mach number of 0.8 (80) with respect to the gas in the lower (upper)
part of the grid. The boundary is perturbed as a sine wave of amplitude 
0.03~kpc. This excites Kelvin--Helmholtz instabilities, with a linear 
growth time of:
\begin{equation}\label{khtime}
\tau_\mathrm{KH}= 5\,\mathrm{Myr} \left(\frac{\lambda}{1\,\mathrm{kpc}}\right)
	\left(\frac{\eta}{10^{-4}}\right)^{-1} \left(\frac{v}{2\times10^7\,\mathrm{m\,s}^{-1}}\right)^{-1}\, ,
\end{equation}
where $\eta$ is the density ratio, and $\lambda$ is the wavelength 
of the perturbation. The perturbations are seeded with a wavelength of 
the box size. However the discretisation seeds perturbations on the scale 
of the resolution, which dominates the evolution. The initial velocity
is in the horizontal direction only, not parallel to the surface.
As the smallest perturbations grow fastest, this speeds up the development towards
turbulence.
The gravitational acceleration is comparatively low, wherefore the linear
Rayleigh--Taylor instability does not contribute significantly to the driving.

Simulation parameters that were varied between simulations are given in 
Table~\ref{simpars}. Note that for the model with the temperature cut in 
the cooling function the cloud density was set to yield the minimum temperature.

\begin{table*}
 \begin{minipage}{140mm}
  \caption{Simulation parameters}\label{simpars}
  \begin{tabular}{@{}rrcccc@{}}
  \hline
  Label & Cloud 	& Total		& Upper 	& Temperature 	& Final\\
  	& density 	& cloud mass	& boundary 	& cut 		& time\\
   	& [$m_\mathrm{p}\,$m$^{-3}$] & [$M_{\rm \mathrm{\sun}}\,$kpc$^{-3}$] & & [K] 	& [Myr]\\
 \hline
ctrl\footnote{contains no clouds, just background density}	
	&   $10^4$	& $1.5\times10^3$	& closed& 0 		& 10 \\
5c0	&   $10^5$		& $1.5\times10^4$	& closed& 0 		& 10 \\
6o0	&   $10^6$		& $1.5\times10^5$	& open	& 0 		& 10 \\
7c4	&   $5\times 10^6$	& $7.4\times10^5$ 	& closed& $10^4$ 	& 10 \\
7o0	&   $10^7$		& $1.5\times10^6$	& open 	& 0 		& 10 \\
7c0	&   $10^7$		& $1.5\times10^6$ 	& closed& 0 		& 6.4 \\
8o0	&   $10^8$		& $1.5\times10^7$ 	& open	& 0 		& 3 \\
 \hline
\end{tabular}
\end{minipage}
\end{table*}

The timestep was found to decrease substantially with contained mass. Hence the high 
mass simulation was stopped after~3~Myr.

\subsection{General dynamics}
We begin by discussing run 7c4 which serves to illustrate the dynamics shown in all 
simulations containing clouds (see Figure~\ref{lgd7c4}). 
At the beginning of the simulations, the dynamics is governed by the 
Kelvin-Helmholtz instability. There is also resonant a kink instability which could 
contribute in principle. This instability produces a characteristic shock pattern 
(a kinked shock) at the interface \citep{BW95}. 
However, we have checked images of the velocity divergence, 
and could find no trace of this pattern. 
The evolving
Kelvin-Helmholtz instability produces vortices in the subsonic lower half
of the computational domain, and
sends strong shocks into the upper medium and the clouds.
Simultaneously, the clouds cool, either to the temperature cut of $10^4$K,
or to about $10^3$K. In this phase, the clouds evolve much like the ones in the shocked cooling
cloud simulations of \citet{MKR02}. Once the shock passes, the clouds are compressed.
\citet{MKR02} then see a fragmentation into small stable cloudlets: 
our simulations do not have sufficient resolution to observe fragmentation. 
Instead we observe the formation of one
filament per cloud. When the contact surface reaches it, the filament spreads along the surface.
The filament is initially unable to penetrate the contact surface due to the long Rayleigh-Taylor time-scale
(Figure~\ref{lgd7c4}, second row, left). Subsequently, one of the clouds is stretched and 
pushed towards the upper boundary by a rising part of the instability.
This particular feature is especially related to the numerical issues detailed 
in section~\ref{num.is}.
Another cloud, located at a declining part of the contact surface is reached by a branch
of the low density gas and drawn into it (Figure~\ref{lgd7c4}, second row, right). 
There it is greatly stretched and soon forms a filamentary system. 
When the first of the dense filaments reaches the lower
boundary (between the centre and the right plots of the bottom row in Figure~\ref{lgd7c4})
the structure of the flow changes. The low density gas can no longer rush through from 
left to right, and the motions turn completely into turbulence. Towards the end,
the upper part moves left and the lower one to the right, forming one large scale vortex.

\subsection{Dynamics of the different setups} 
\begin{figure}
\vspace*{3cm}
\centering
\includegraphics[width=0.48\textwidth]{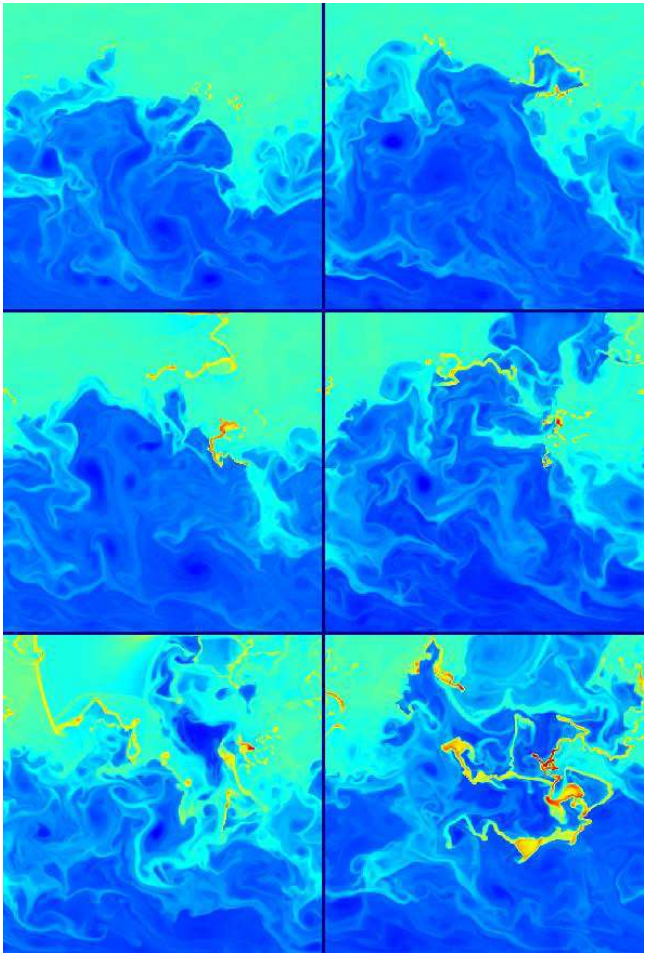}
\caption{Logarithmic density plots at 3 Myr for all the simulations with clouds in the order of 
	Table~\ref{simpars} (5c0: top left, 6o0: top right, 7c4: middle left,
	7o0: middle right, 7c0 bottom left, 8o0: bottom right).
	An open upper boundary (right plots) amplifies the Kelvin-Helmholtz instability.
	More cold mass (towards the bottom) speeds up the formation of filaments.}
\label{lgd3myr}
\end{figure}
Logarithmic density plots at 3~Myr are shown in Figure~\ref{lgd3myr} for all 
six simulations containing clouds. The mass loading clearly affects the evolution
of the Kelvin-Helmholtz instability. 
This is most evident when comparing the right column
of Figure~\ref{lgd3myr} (open upper boundary). 
When the first filament is stretched towards the upper boundary, the low density gas 
flows along, leaving the grid if the upper boundary is open. In the simulations with lower
cloud density,
this feature is suppressed, and the low density gas stays in the lower part
of the computational domain. Also for the simulations with lower initial cloud mass 
the overall evolution is slower. 

The direct comparison 
between open and closed boundary (runs 7c0 and 7o0) shows that dense gas assembles near
the boundary for the closed simulations and that this gas leaves the grid for the open ones.
This is not compensated by inflows, and hence, the open boundary cases have less
cold mass than the closed upper boundary simulations.

The temperature cut results in a significantly reduced maximum density (by a factor
of order $10$). The densest cloud occupies a bigger area, i.e. it is less condensed.
Also, the aforementioned uplifting of low density gas is suppressed.

\subsection{Power spectrum}
\begin{figure}
\centering
\includegraphics[width=0.23\textwidth]{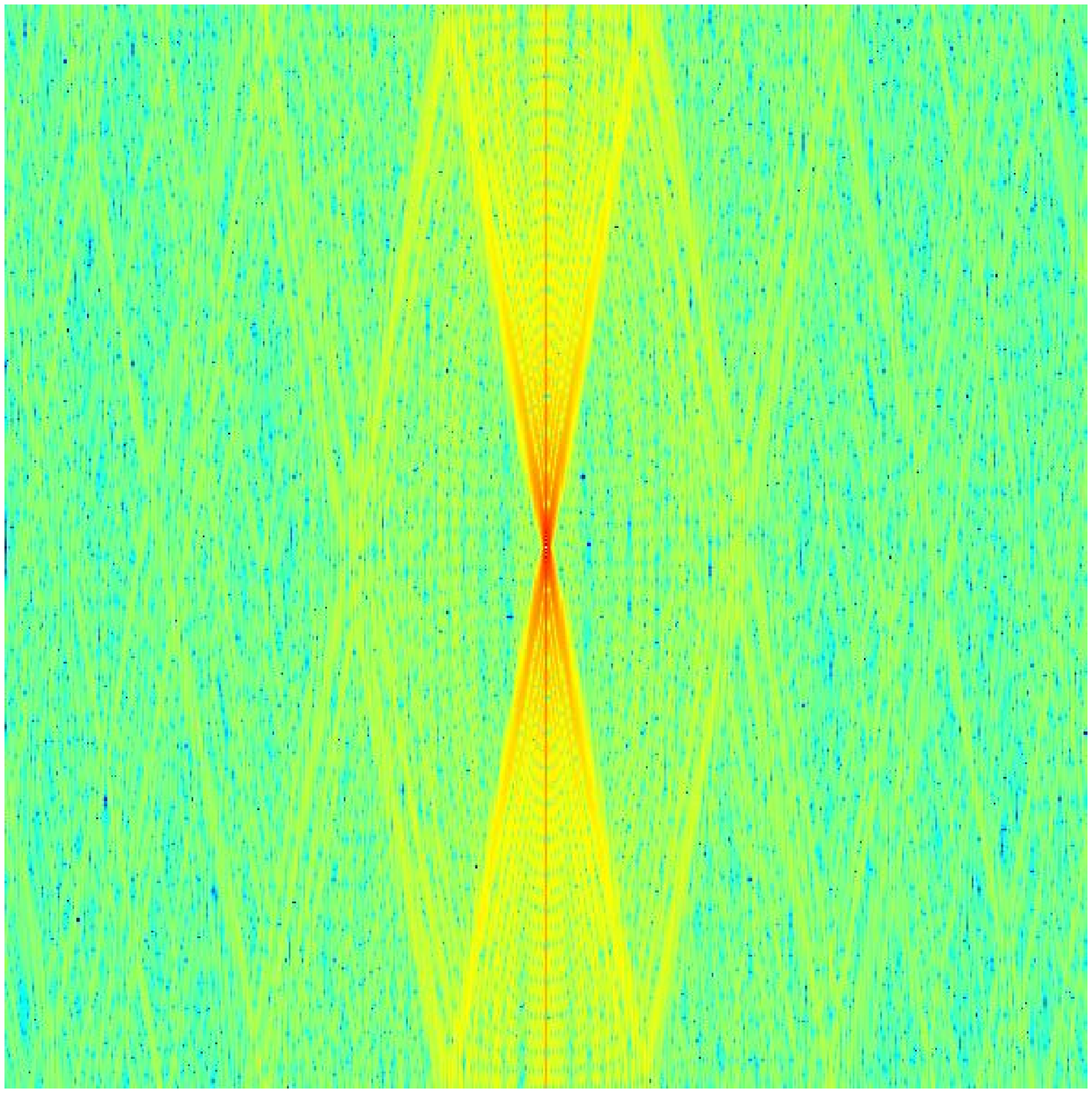}
\includegraphics[width=0.23\textwidth]{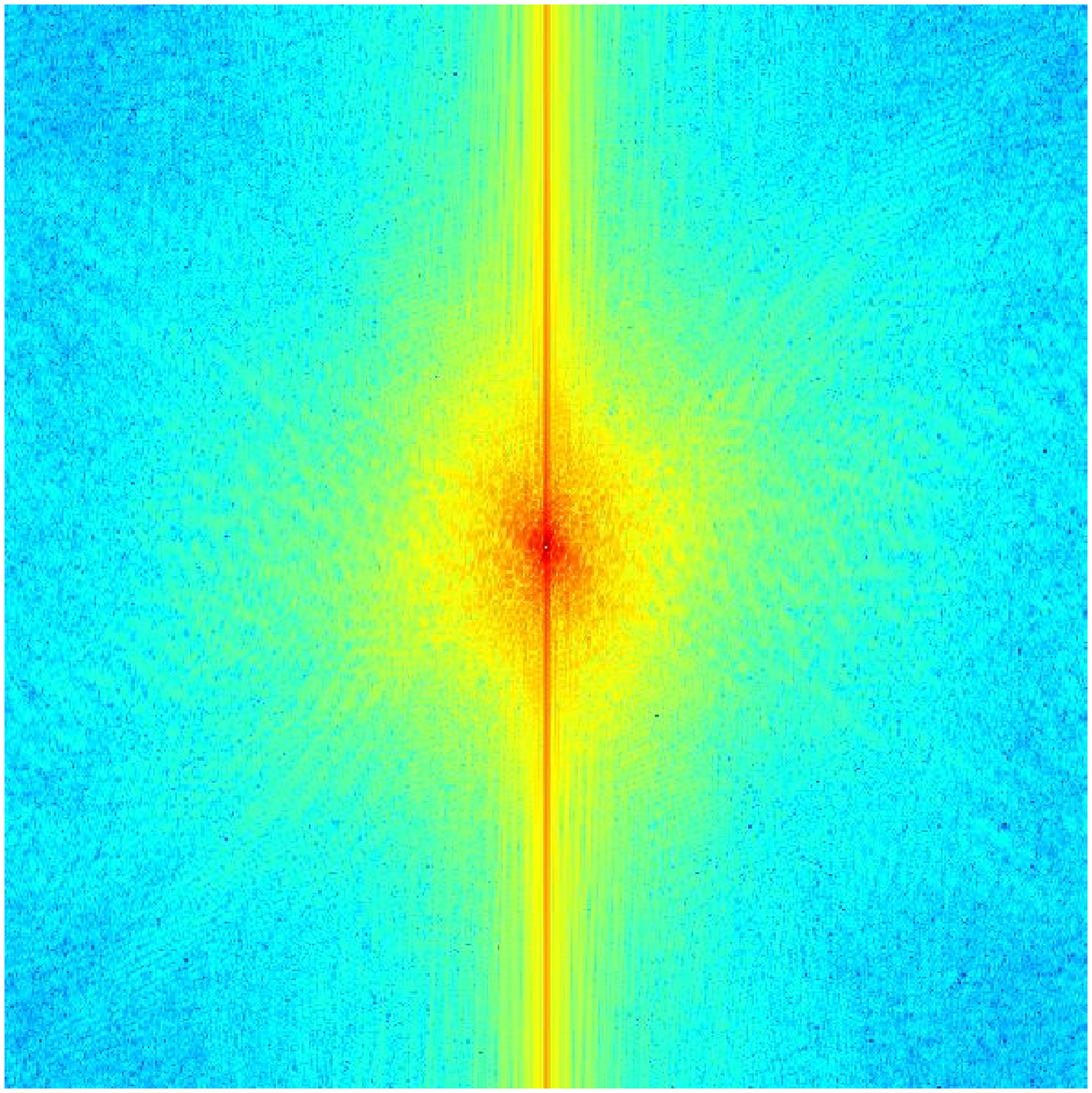}
\includegraphics[width=0.23\textwidth]{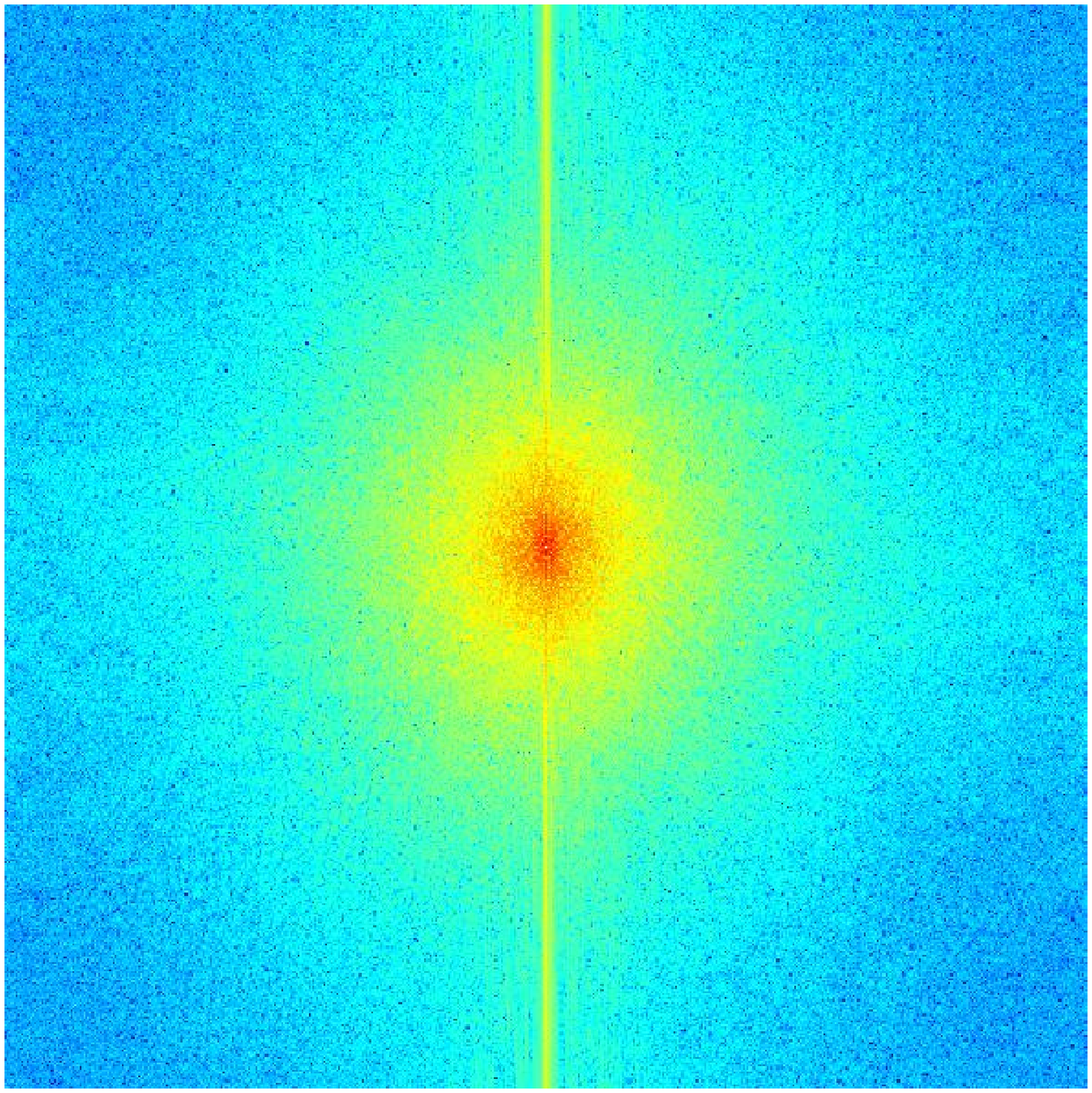}
\includegraphics[width=0.23\textwidth]{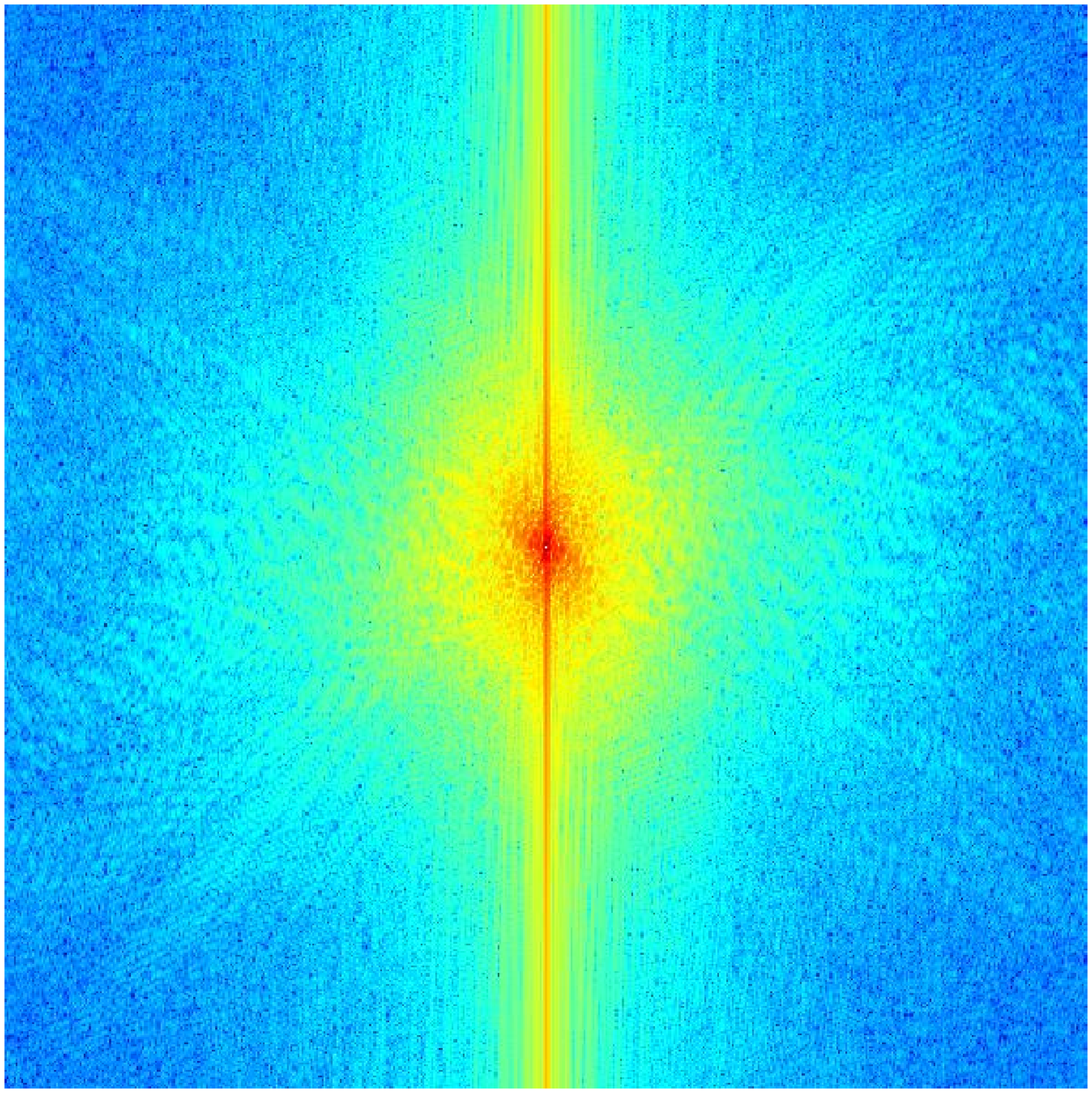}
\caption{Two-dimensional velocity power spectra 
	for run 7c4 at 0.2~Myr (top left),
	3.2~Myr (top right), and 7.7~Myr (bottom left) and for run~7o0
	at $t=7.7$~Myr (bottom right). The power spectra are evaluated 
	in the zero momentum frame. Wave-vector zero is at the centre.}
\label{v2pow2d}
\end{figure}
\begin{figure}
\centering
\includegraphics[height=0.35\textwidth, angle=0]{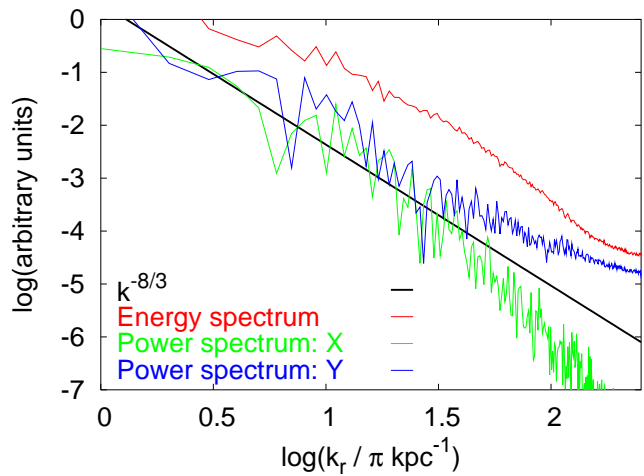}
\caption{Energy spectrum ($E(k_r) dk_r=\int_{k_\phi} P(k_r, k_\phi) k_r dk_\phi dk_r$), 
	X-~and~Y-cuts through the power spectrum of run 7c4 at 7.7~Myr.
	The large scales isotropise first, showing a Kolmogorov-like decline 
	($k^{-8/3}$ in the power spectrum cuts, corresponding to $k^{-5/3}$ in the 
	energy spectrum). The smallest scales do not isotropise up to the end of the
	simulation, which is due to the bar feature.}
\label{v2pow1d}
\end{figure}
For comparison to turbulence studies, it is important to check isotropy and 
scale-dependence. Therefore, we examined the power spectrum of the velocity,
\begin{displaymath}
P({\bf k})=\sum_{i=x,y}|\int_\mathrm{box} v_i e^{\imath{\bf k x}} d^2 x|^2.
\end{displaymath}
We use velocities in the zero momentum frame although the results do not 
depend on that particular choice.
Two-dimensional representations of the power spectrum are shown in Figure~\ref{v2pow2d}.
The behaviour for the different runs is very similar. Hence, only results from 
runs~7c4, and~7o0 are shown. Even the early power spectra look quite isotropic,
apart from a horizontal bar feature.
However, the energy spectra up to a few million years are dominated by the bar feature,
which produces power law indices around $-2$. Large scales isotropise first,
and at 7.7~Myr, we find a Kolmogorov-like isotropic power spectrum for most scales
(Figure~\ref{v2pow1d}).
The bar feature varies in strength, but is generally decreasing. 
The open upper boundary simulation showed a stronger feature at the same time.
The same is true for simulations with less cold mass load. 

We conclude that for larger scales, isotropic, Kolmogorov-like turbulence is 
essentially reached
after a few million years of simulation time. Some anisotropy remains on small scales.
 
\subsection{Mach number, density and pressure distributions}
\begin{figure}
\centering
\includegraphics[height=0.45\textwidth, angle=-90]{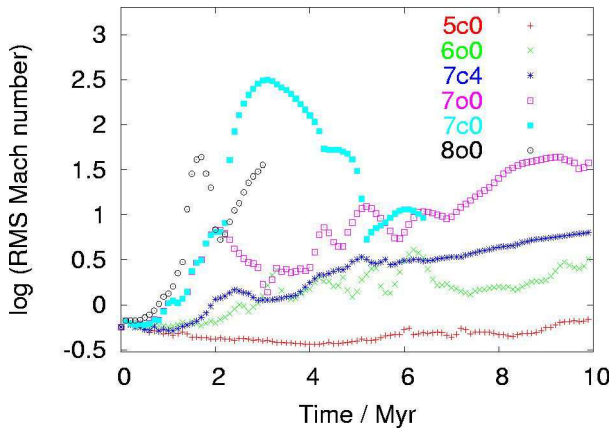}
\caption{Root mean squared Mach number versus time for all runs with clouds. 
	The Mach number generally grows with time. 
	The simulations with higher mass load have higher average Mach numbers.
	The high amplitude features are due high density filaments hitting a boundary,
	and are especially pronounced for closed boundaries..
}
\label{mach.rms}
\end{figure}
The root mean squared (rms) Mach number is shown in Figure~\ref{mach.rms}.
It is generally well above one with the exception of run~5c0.

The choice of boundary condition has a big impact on the rms 
Mach number. We compare the three simulations with roughly the same initial density
(7o0,7c0,7c4). The big bump in the 7c0 line is produced by one large filament hitting
the upper boundary. The latter releases so much kinetic energy that the rms Mach number 
reaches up to a factor of ten larger than in the open boundary case for the same density. 
Remarkably, after relaxation, the two simulations rejoin the same path. 
The bump is completely damped in the simulation with the temperature cut (7c4). 
The temperature cut reduces the peak values of the density, which leads to the more modest interaction with the 
grid boundary.

The most evident feature in Figure~\ref{mach.rms} is the constant rising of the rms 
Mach number and the dependence on the mass loading. More cold mass results in higher
rms Mach numbers. This observation can be explained by a remarkable relationship
between the density and the Mach number. 

\begin{figure*}
\centering
\includegraphics[width=0.49\textwidth]{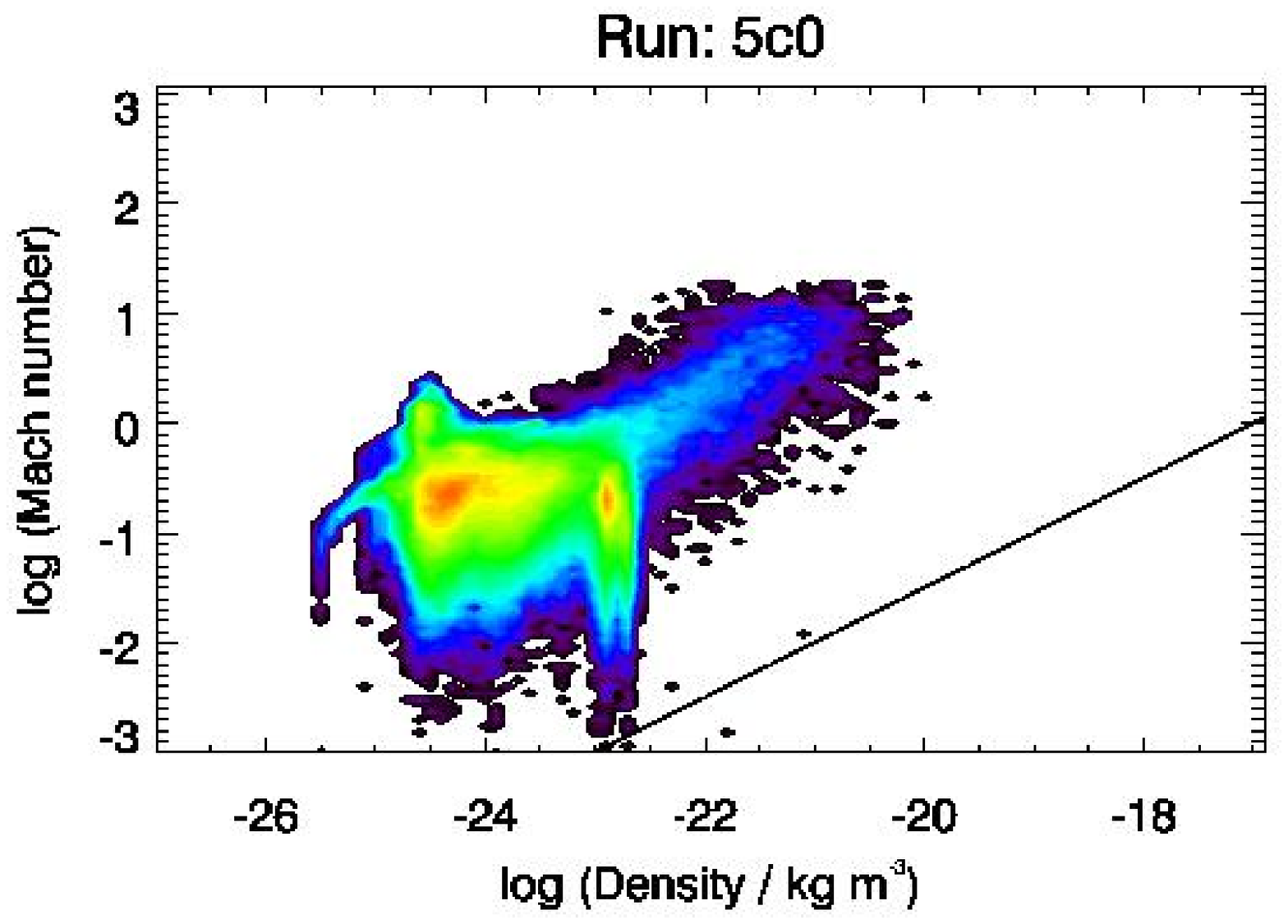}
\includegraphics[width=0.49\textwidth]{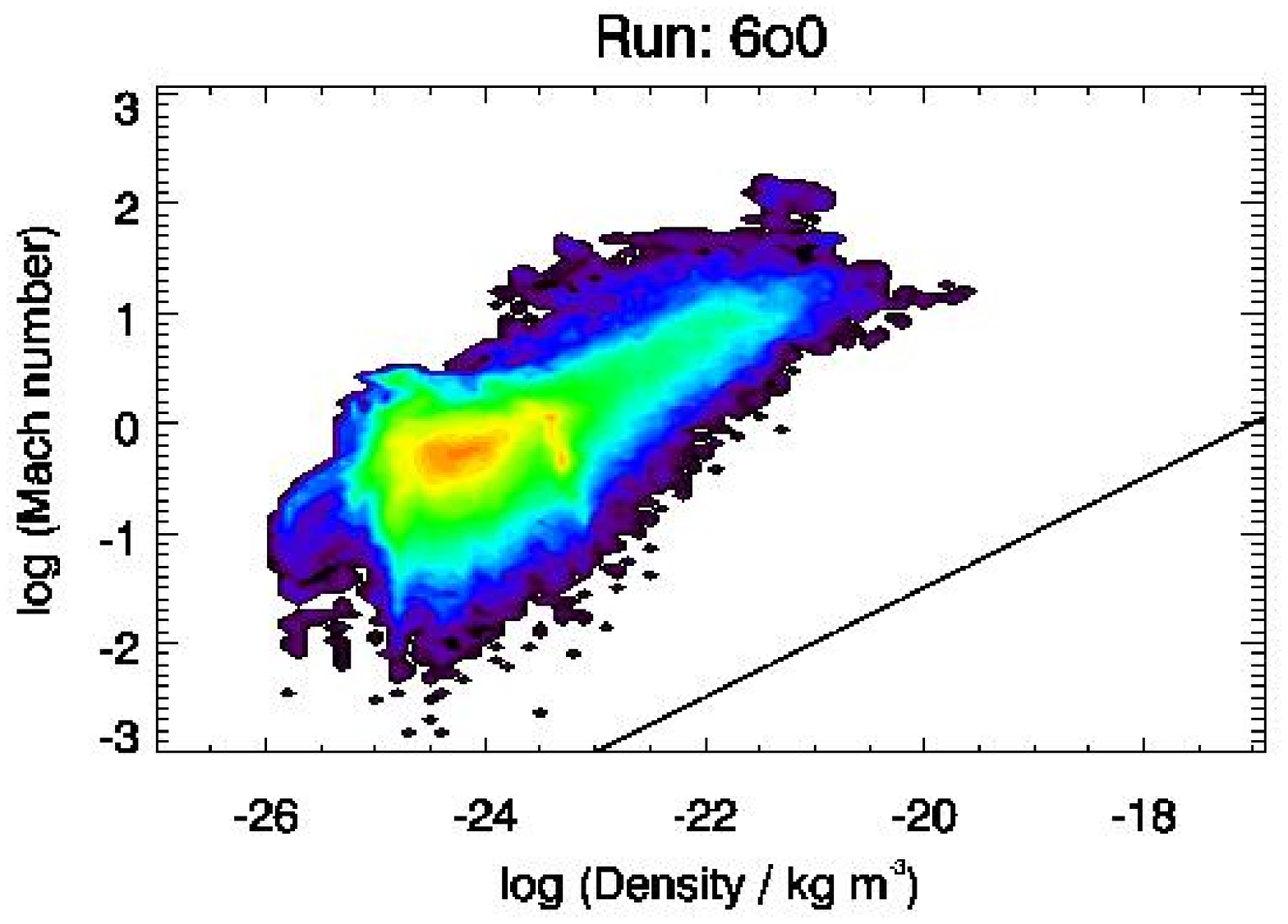}
\includegraphics[width=0.49\textwidth]{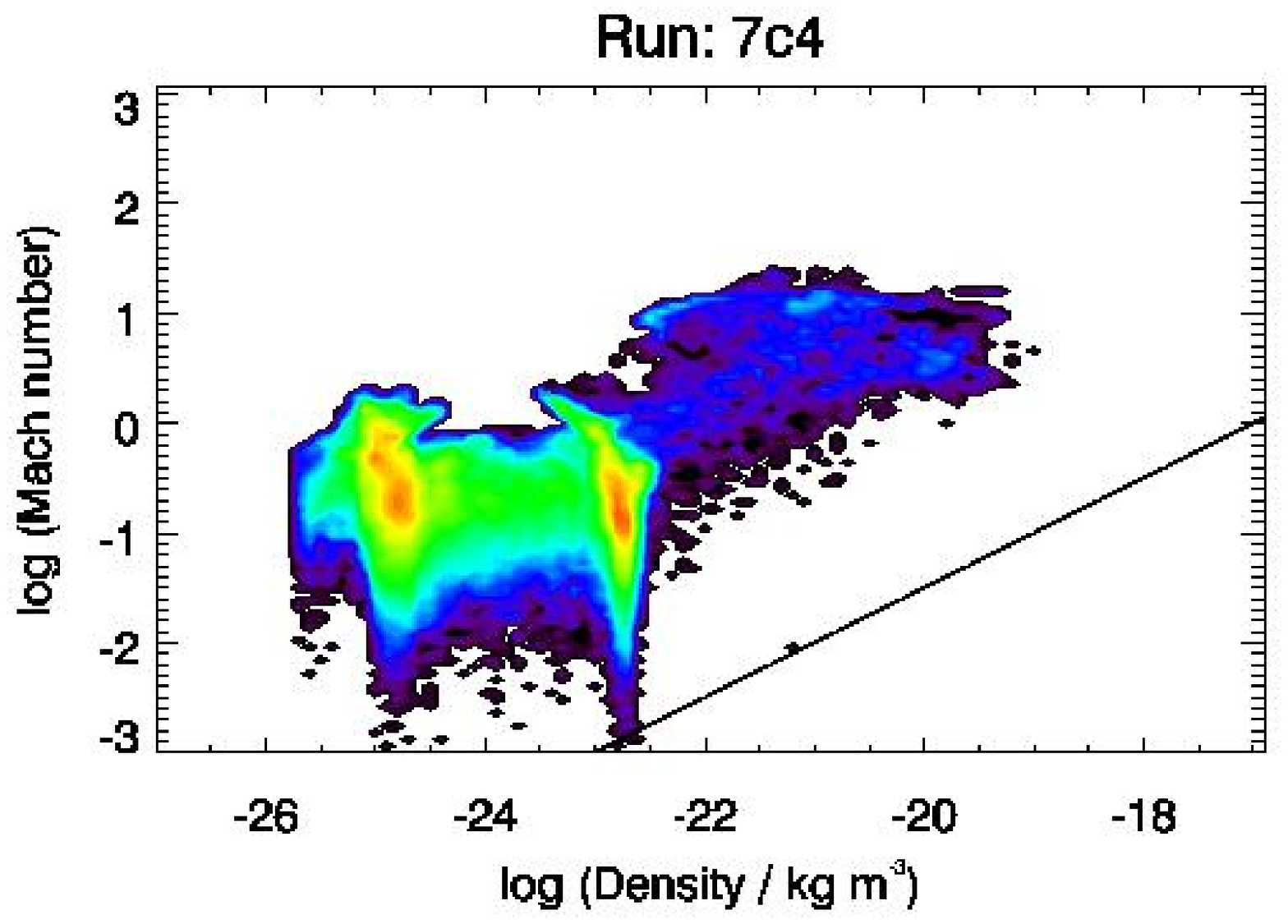}
\includegraphics[width=0.49\textwidth]{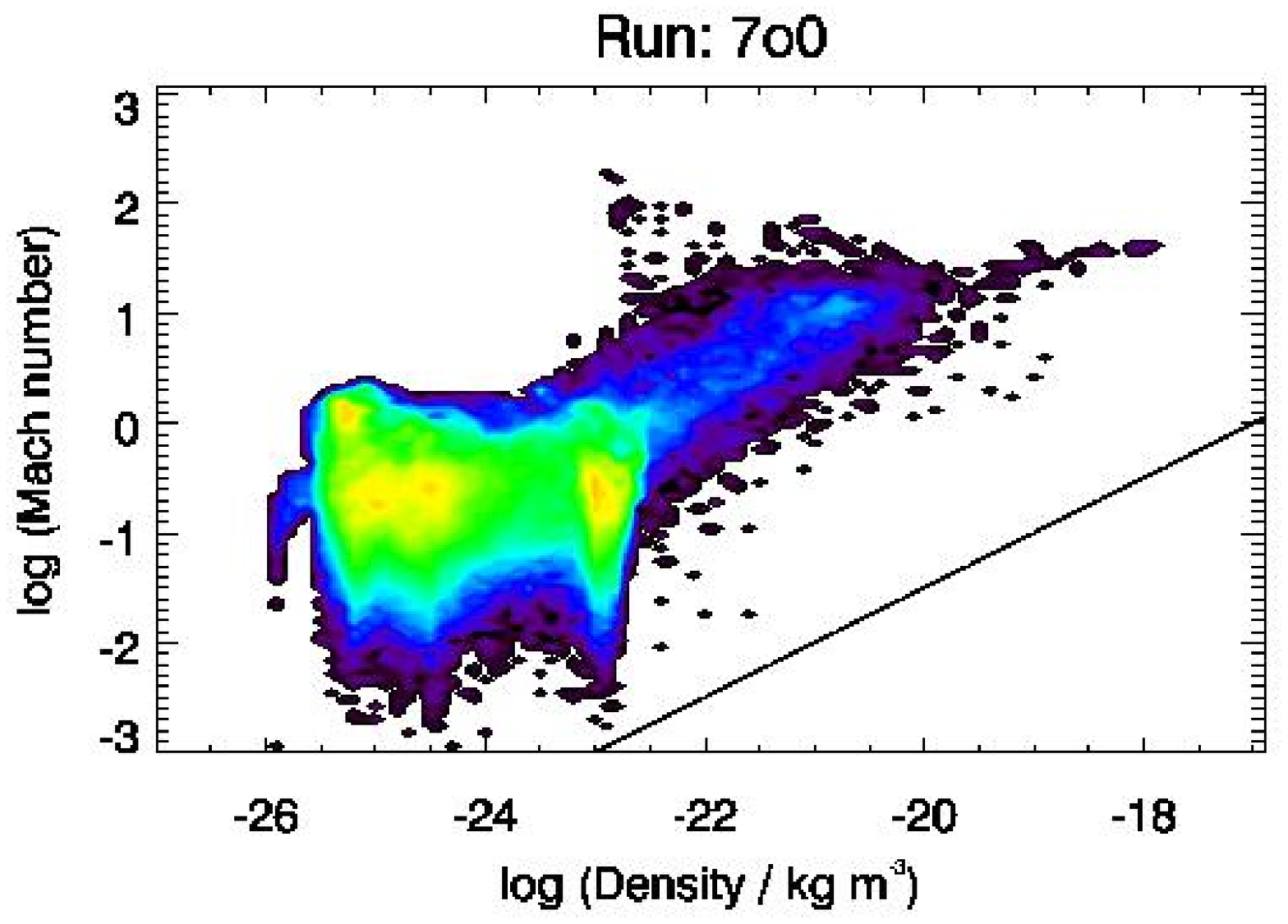}
\caption{Phase diagrams showing the logarithmic volume occupation at a given 
	density and Mach number for runs 5c0 and 6o0 (top at 10~Myr), 
	and runs 7c4 and 7c0 (bottom at 3~Myr).
	The distribution is bimodal: low density gas is unable to cool and forms 
	a turbulent system where the Mach number distribution is independent of 
	the density. Gas with short cooling times joins at a density of about 
	$10^{-23}$~kg$\,$m$^{-3}$ and its Mach number grows roughly with the square 
	root of the density. The fiducial line indicates a $M \propto \rho^{1/2}$ 
	dependence. Colour coding is the same as on other figures.}
\label{dmh}
\end{figure*}

The relationship between density and Mach number is shown in the 
2D-histograms of Figure~\ref{dmh}. These diagrams display the volume occupation
of gas at a given density and Mach number. We have chosen the time of 3~Myr 
(10~Myr) for the higher (lower) cold mass load simulations, because the turbulence 
is well evolved at that time, with the mixing between the low and medium density 
gas still being acceptably small. These plots look very similar at most other 
simulation times. Mixing of these gas phases is indicated by the two peaks that are 
initially at $10^{-27}$~kg$\,$m$^{-3}$ and $10^{-23}$~kg$\,$m$^{-3}$ coming closer 
together.

\begin{figure*}
\centering
\includegraphics[width=0.49\textwidth]{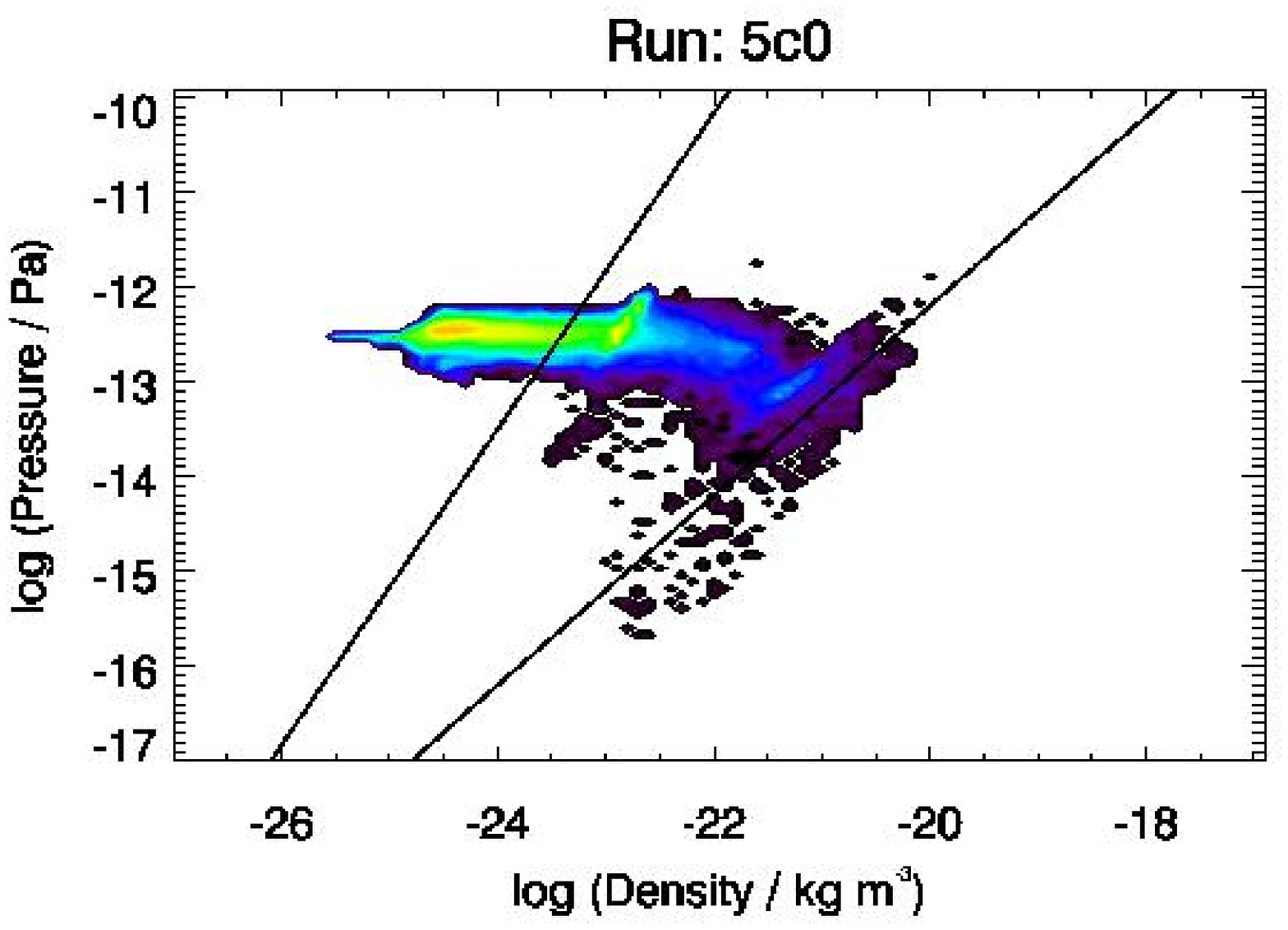}
\includegraphics[width=0.49\textwidth]{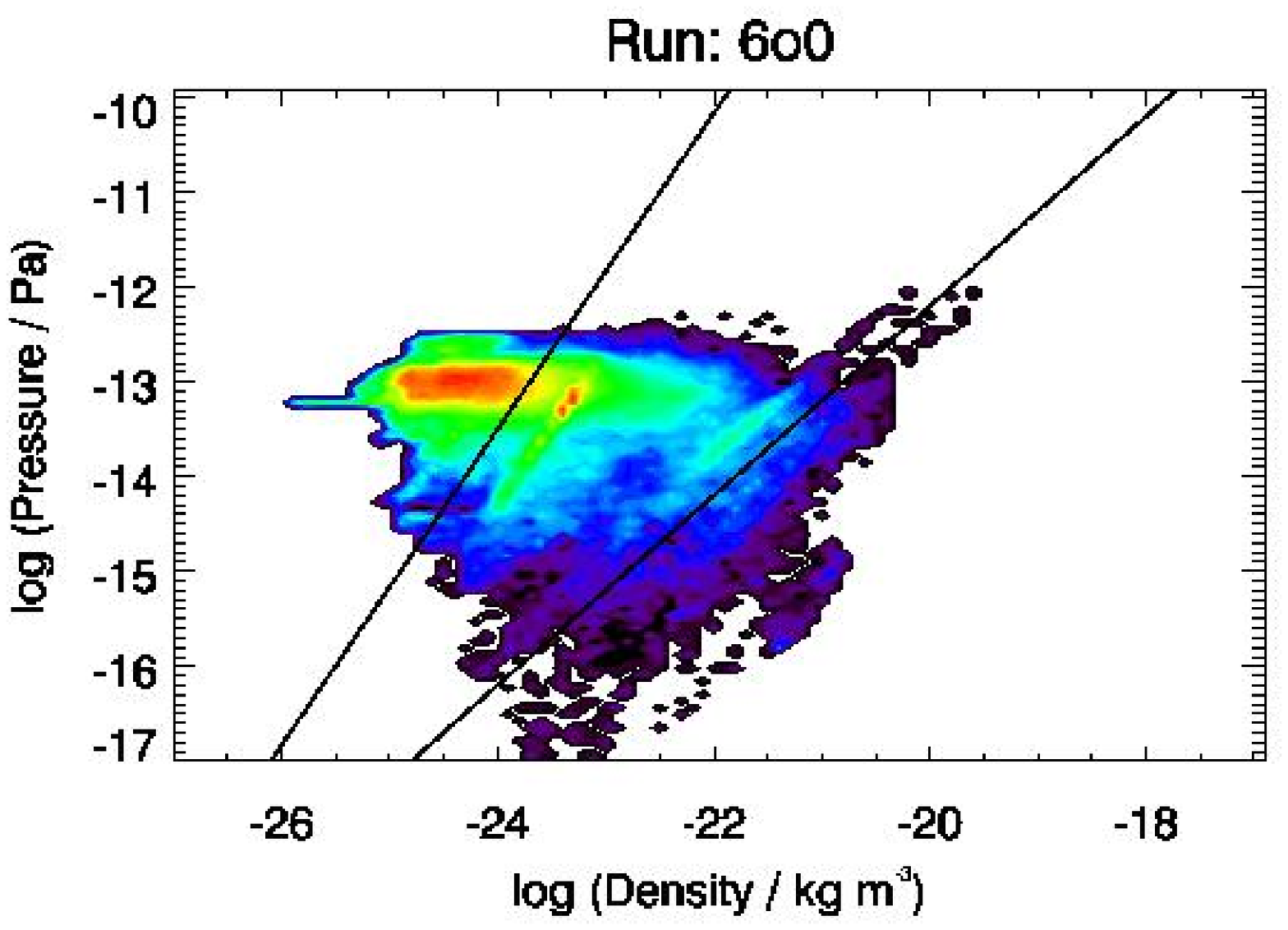}
\includegraphics[width=0.49\textwidth]{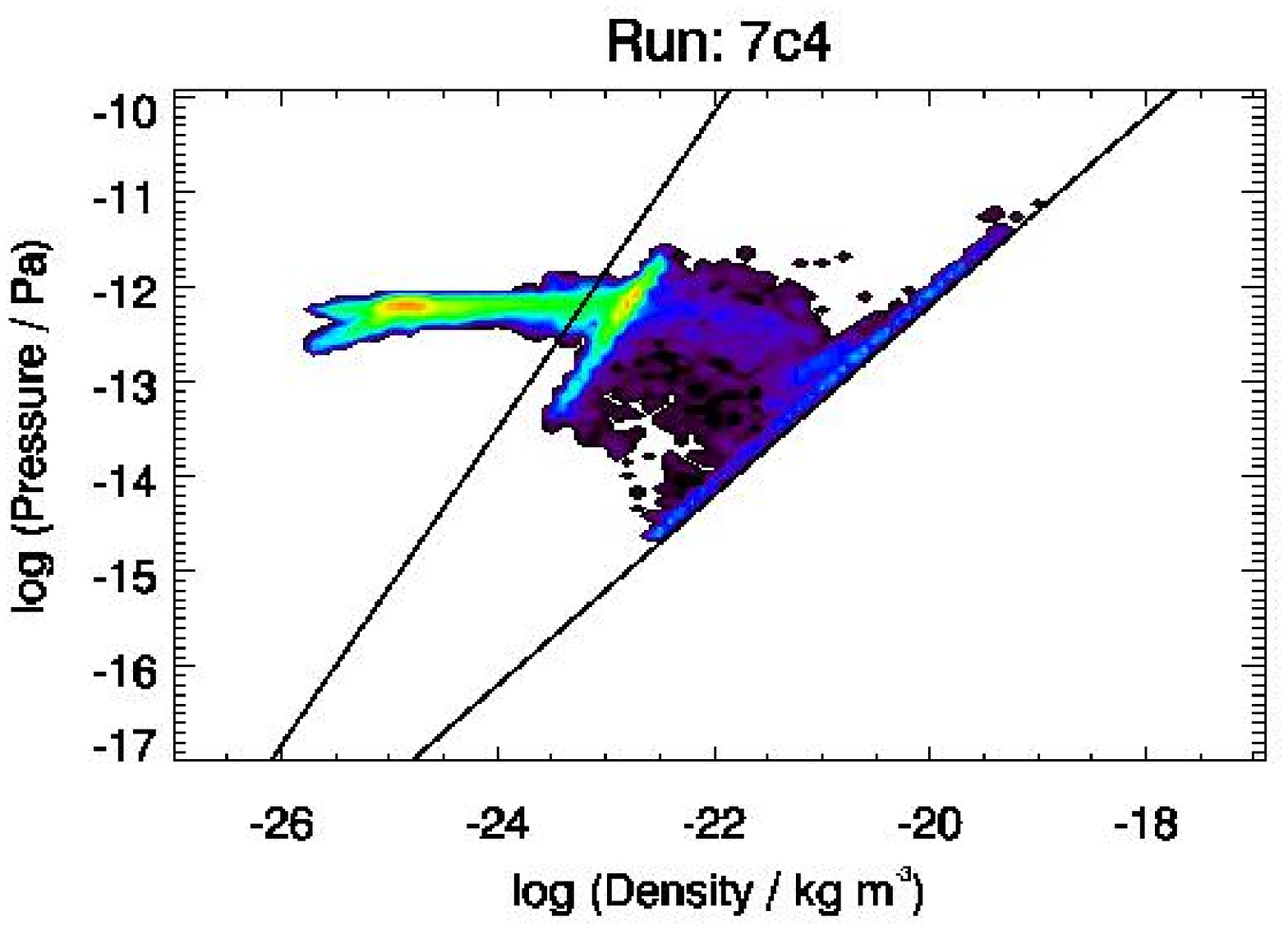}
\includegraphics[width=0.49\textwidth]{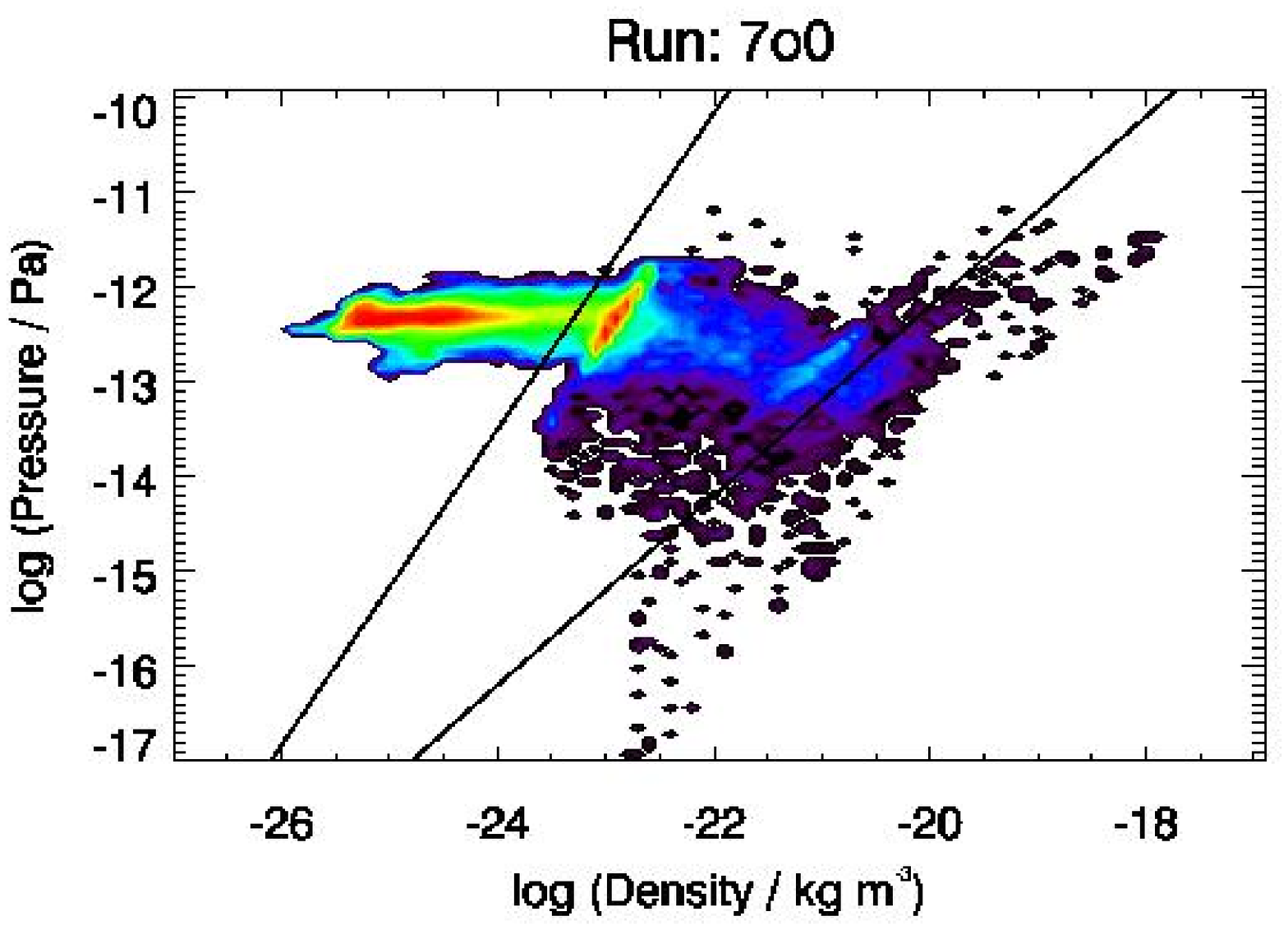}
\caption{Phase diagrams showing the logarithmic volume occupation at a given 
	density and pressure for runs 5c0 and 6o0 (top at 10~Myr), 
	and runs 7c4 and 7c0 (bottom at 3~Myr). The pressure of much of the 
	simulated volume is within a factor of a few, even in the strongly 
	radiating part. Fiducial lines of slope $5/3$ and~1 highlight  
	adiabatic expansion features, prominently around a density of 
	$10^{-23}$~kg$\,$m$^{-3}$ (upper background gas), and isothermal 
	components, respectively. In run 7c4, the $10^4$~K line marks the border 
	of the accessible phase space, in the other simulations,
	gas assembles at a slightly higher temperature due to the drop of the 
	cooling function. Colour coding is the same as on other figures.}
\label{dph}
\end{figure*}

Between these densities, the Mach number is independent of the density. Since the 
simulation is effectively isobaric at those densities (see Figure~\ref{dph}), this 
means that the kinetic energy density is roughly constant, indicative of a 
quasi-equilibrium state. The same relation would be expected in a shock dominated 
scenario, where $T\propto v^2$. At high densities, the internal energy is reduced 
by the radiation losses, which increases the Mach number. Although this can change 
the pressure by many orders of magnitude in individual cells, the bulk of the 
material can compensate for the pressure loss by contraction and shock heating, 
which brings the gas back to near isobaric conditions. Since the velocity is not 
correlated with the density, the dependence of the Mach number on density 
($M^2=\rho v^2/\gamma p$) is dominated by the explicit density dependence.
These findings agree well with the 3D simulations of \citet{KN04} who drive the 
turbulence by the thermal instability, only, and do not include the low density 
phase.

The $M\propto \rho^{1/2}$ branch has got a higher occupation and extends to higher
Mach number for higher mass loading and later simulation times. This causes the 
behaviour of the rms~Mach number in Figure~\ref{mach.rms}.

\begin{figure*}
\centering
\includegraphics[width=0.49\textwidth]{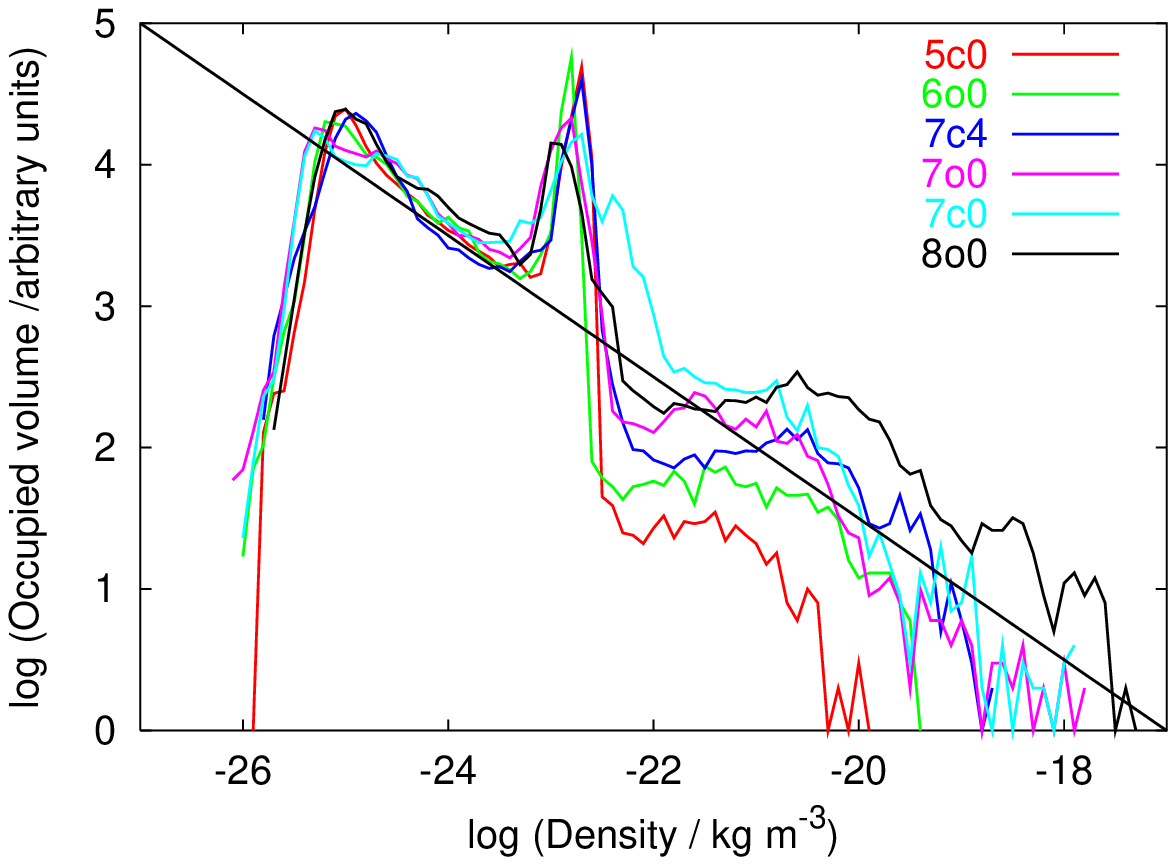}
\includegraphics[width=0.49\textwidth]{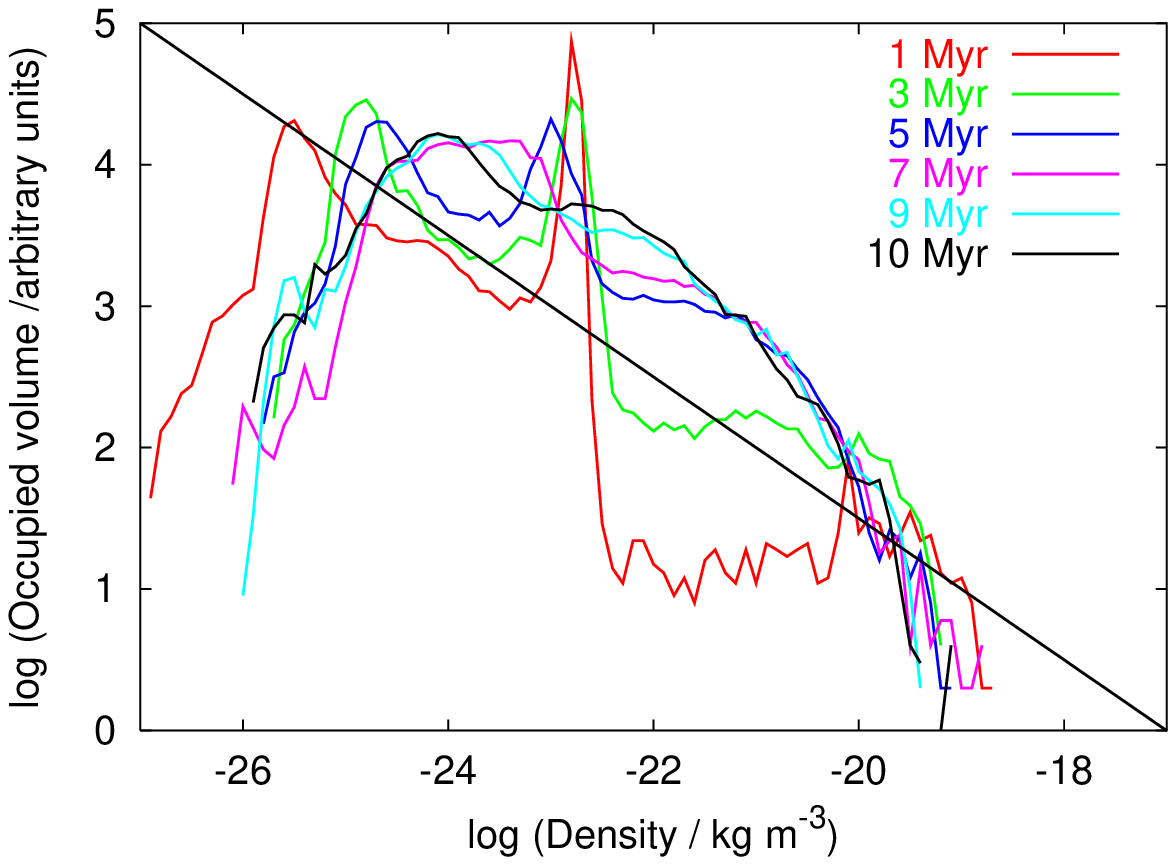}
\caption{Left: Density probability distribution functions at a representative time of 2.5~Myr for all
	simulations. The fiducial line is proportional to $\rho^{-1/2}$.
	Right: The same for run 7c4 for a series of simulation times.}
\label{lgd-pdf}
\end{figure*}
 
Typical pressure versus density histograms are shown in Figure~\ref{dph}. Although the 
occupied phase space spans many orders of magnitude in both dimensions, the pressure
is remarkably peaked towards a central value, even in regions with high density and a shorter 
cooling time -- here the pressure is, in general, at most a factor of ten below the 
low density gas. The prominent linear features in these figures are due to two processes.
One is quasi-adiabatic expansion and compression of the marginally 
cooling gas at $10^{-23}$~kg$\,$m$^{-3}$, which initially fills the most 
of the upper part of the grid. The other is the quasi-isothermal regime at the 
prominent drop of the cooling function at about $10^4$~K. The line is most prominent
in the simulation with the cut in the cooling function but is also apparent in the other
simulations at a temperature of about 14,000~K. Evidently, at this point the energy gain due to shocks
caused by the turbulent motions balance the radiation losses.

The density PDFs are dominated by a power law behaviour (Figure~\ref{lgd-pdf}).
As long as the peaks of the initial condition remain visible, the region between the peaks
tends to follow a $V(\rho) \propto \rho^{-1/2}$ law, where $V(\rho)$ is the occupied volume 
at a particular density $\rho$. The high density part of the distribution reflects the mass loading,
and is independent of the low density part. This region also follows the $\rho^{-1/2}$ law,
at least for higher mass load. Over time, the region around $10^{-22}$~kg$\,$m$^{-3}$
becomes more and more populated. As the peaks diffuse, the distribution becomes uniform 
with exponential type cutoffs towards low and high density.

\subsection{Temperature distribution}\label{tdistsec}
\begin{figure*}
\centering
\includegraphics[width=0.49\textwidth]{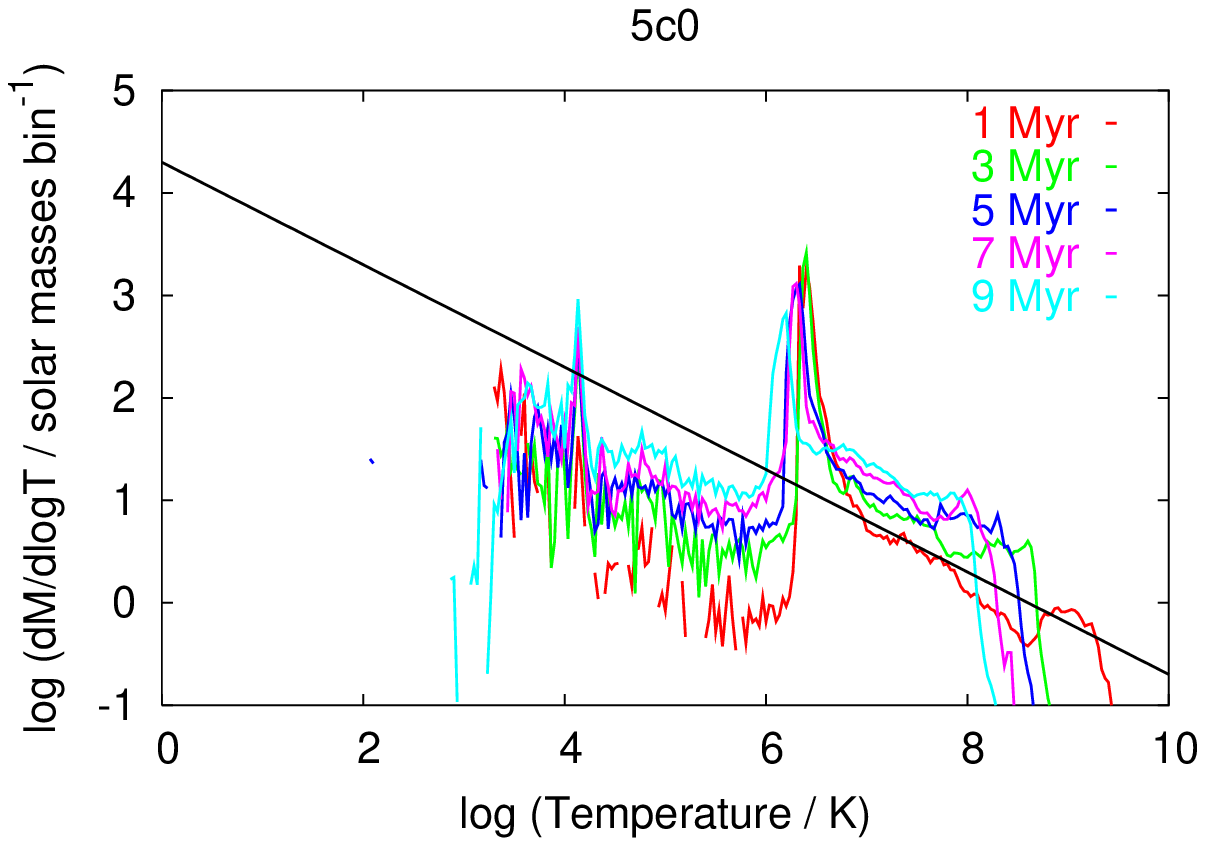}
\includegraphics[width=0.49\textwidth]{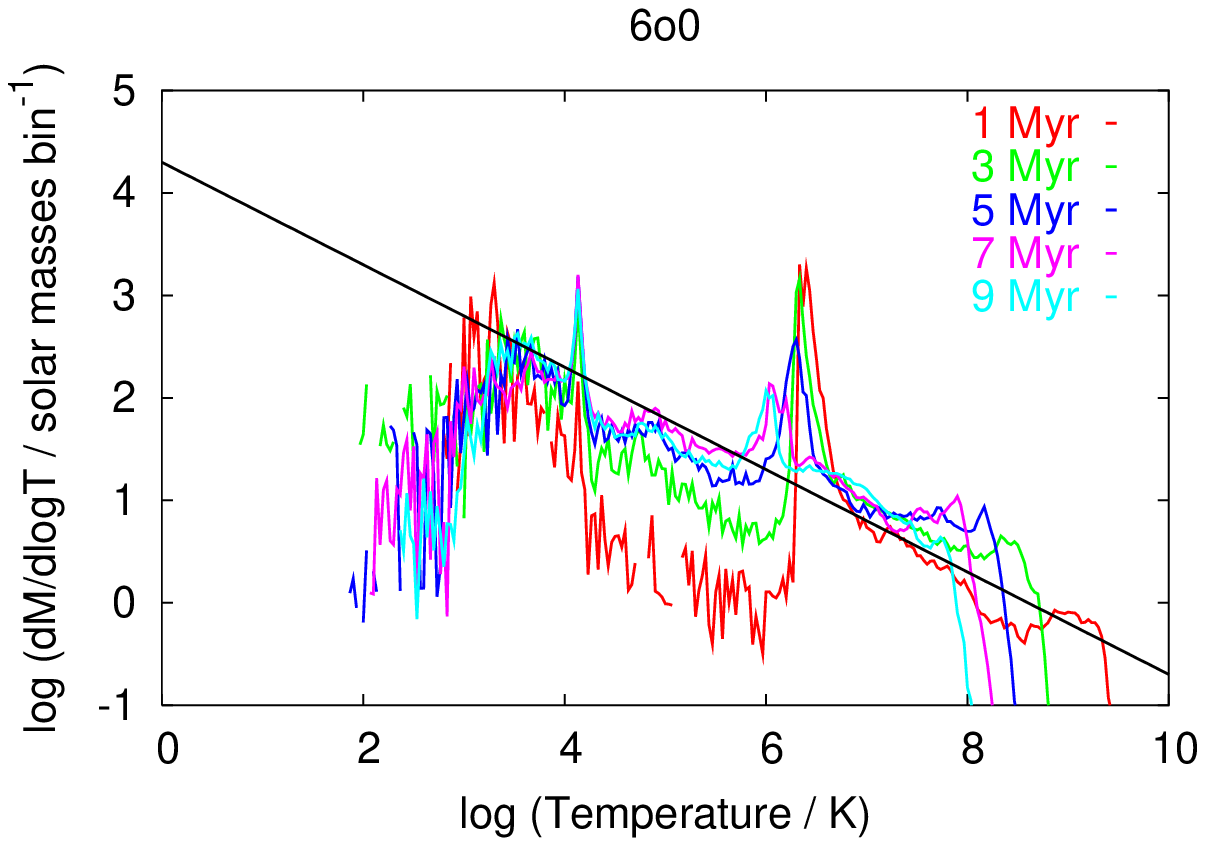}
\caption{Gas mass distribution versus temperature for runs 5c0 and 6o0 
 	at different 
	simulation times. The black line is proportional to $1/\sqrt{T}$. 
	The size of the temperature bins was 0.03.
}
\label{tmasslog}
\end{figure*}
We show the distribution of gas mass as a function of temperature in 
Figure~\ref{tmasslog}. The general behaviour is well described by a 
$dM/dT \propto 1/\sqrt{T}$ law. The high temperature part ($T>10^6$~K) evolves independently 
from the cooler part.
Due to numerical mixing, the upper temperature of the gas moves to lower temperature 
keeping an exponential-type cutoff. In all simulations, the distribution evolves quickly
to follow a $dM/dT \propto 1/\sqrt{T}$ form for $T> 10^6$~K up to the cutoff.
There is a well defined peak at $T \approx 10^6$~K which broadens during the simulation.
For $T<10^6$~K the distribution evolves more slowly towards the same 
$dM/dT \propto 1/\sqrt{T}$ established at higher temperatures and at a rate which increases
with initial cloud mass. A peak at 14,000~K is well-defined in all simulations, 
and does not evolve with time. Below $\approx 10^3$~K the distribution evolves to form what may be
a power-law decline at low temperatures.

\subsection{Filling factor and mass}\label{ffm}
\begin{figure*}
\centering
\includegraphics[width=0.49\textwidth]{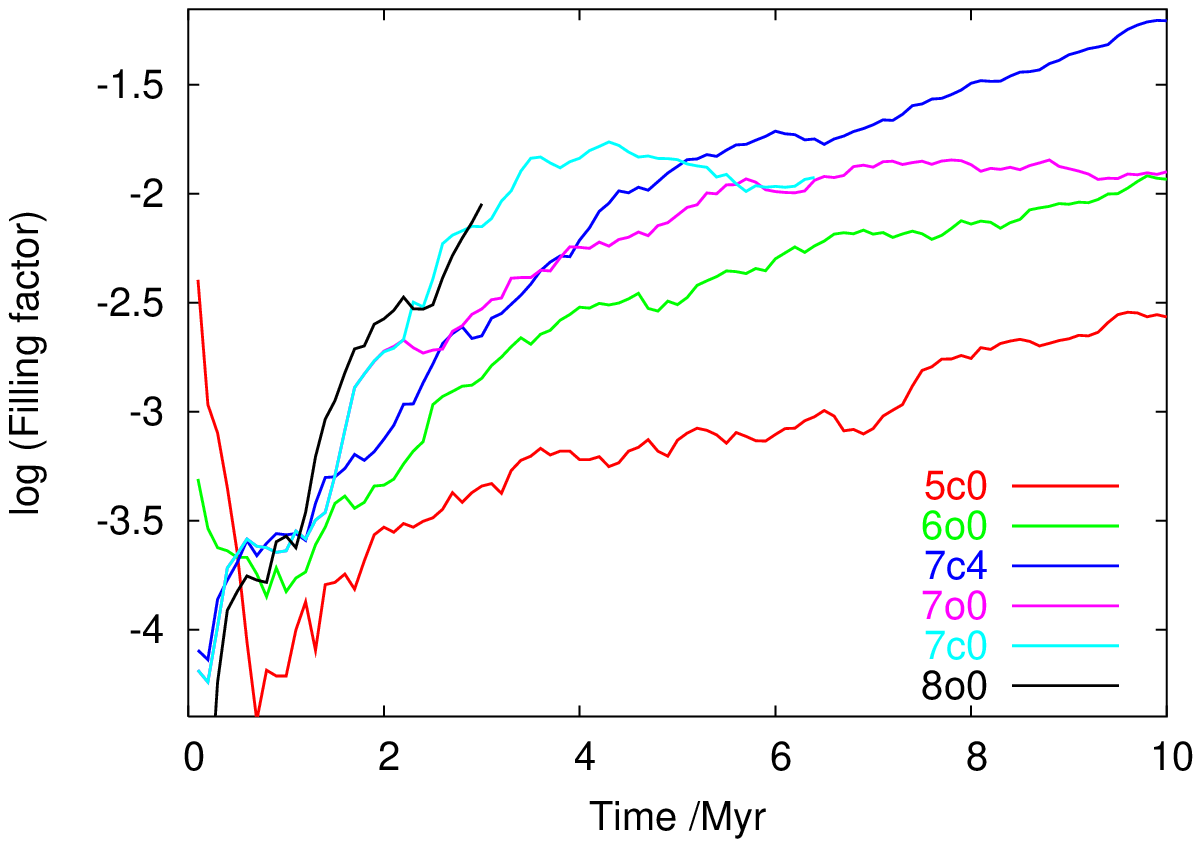}
\includegraphics[width=0.49\textwidth]{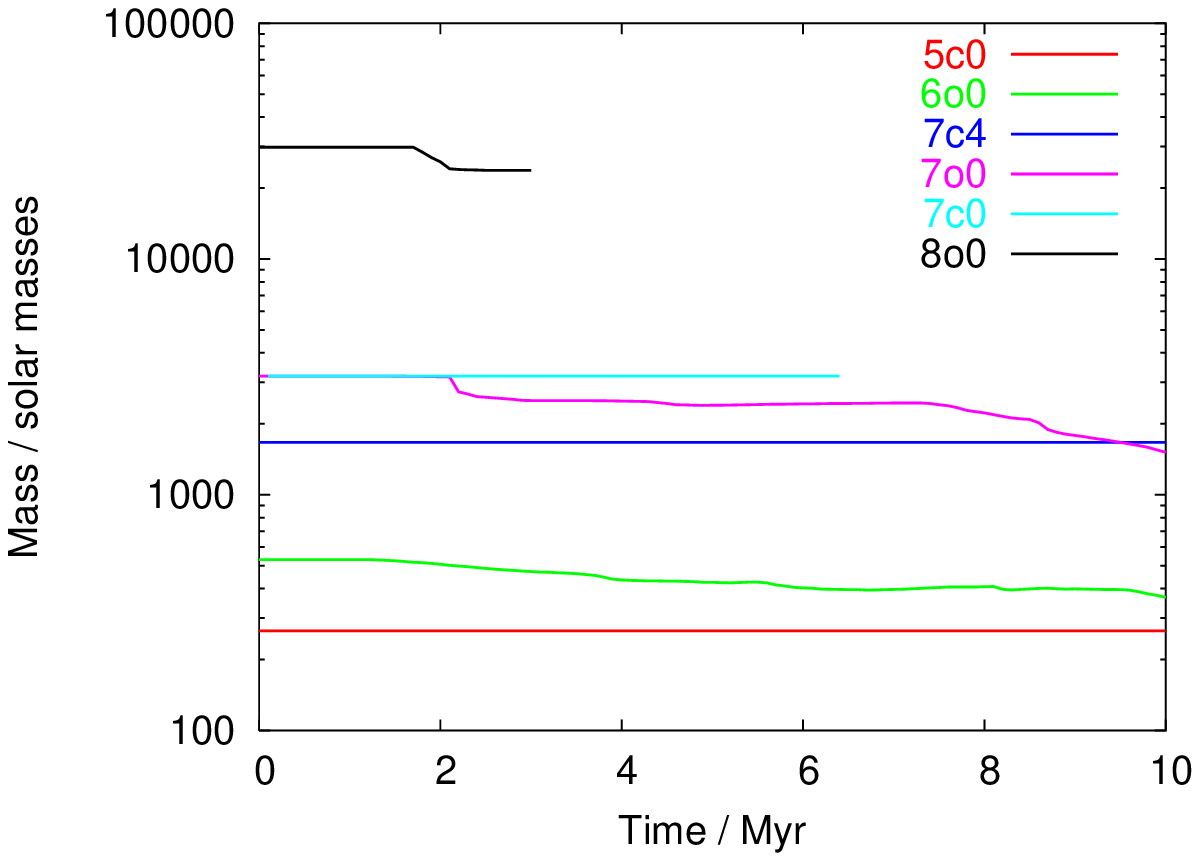}
\caption{Filling factor (left) of gas in the 14,000~K peak and total mass (right) 
over time for all simulations with clouds. The mass is conserved by NIRVANA, and therefore, 
only the simulations with open boundary lose some mass (up to $\approx 50\%$)}
\label{ff+mass}
\end{figure*}
NIRVANA conserves the total mass to machine accuracy. In the simulations with closed boundaries,
the mass is therefore constant (Figure~\ref{ff+mass}, right). The other simulations lose up to about
50 per cent of the mass over the grid boundary -- the ordering according to initial mass loading 
is essentially maintained throughout the simulation. 

Since radiation transfer is not included into the simulation, the ionisation 
structure and nebular emission properties cannot be deduced in detail. However, from
the strong drop of the gas mass towards higher temperature (Figure~\ref{tmasslog}), 
and the drop of the cooling function towards lower temperature, it is clear that 
the gas with a temperature close to 14,000~K is responsible for much of the optical 
luminosity. In Figure~\ref{ff+mass} (left) we show the volume filling factor of gas 
with temperature 14,000$\pm$2,000~K, which corresponds to the peak in the mass 
versus temperature distribution in Figure~\ref{tmasslog}. This filling factor 
depends mostly on mass, with the three most accurate simulations
(5c0, 6o0, and 7c4, see Section~\ref{num.is} below) following nearly parallel paths. 
As expected, higher initial mass loading results in 
a higher filling factor. For most of the simulation times, the filling factor stays 
between $\approx 10^{-3}$ and $\approx 10^{-2}$. Since these are 2D simulations, 
this would correspond to a volume filling factor of between $10^{-5}$ and $10^{-3}$ 
in 3D, assuming similar structuring in all dimensions.

\subsection{Energy loss}\label{energyloss}
\begin{figure*}
\centering
\includegraphics[width=0.49\textwidth]{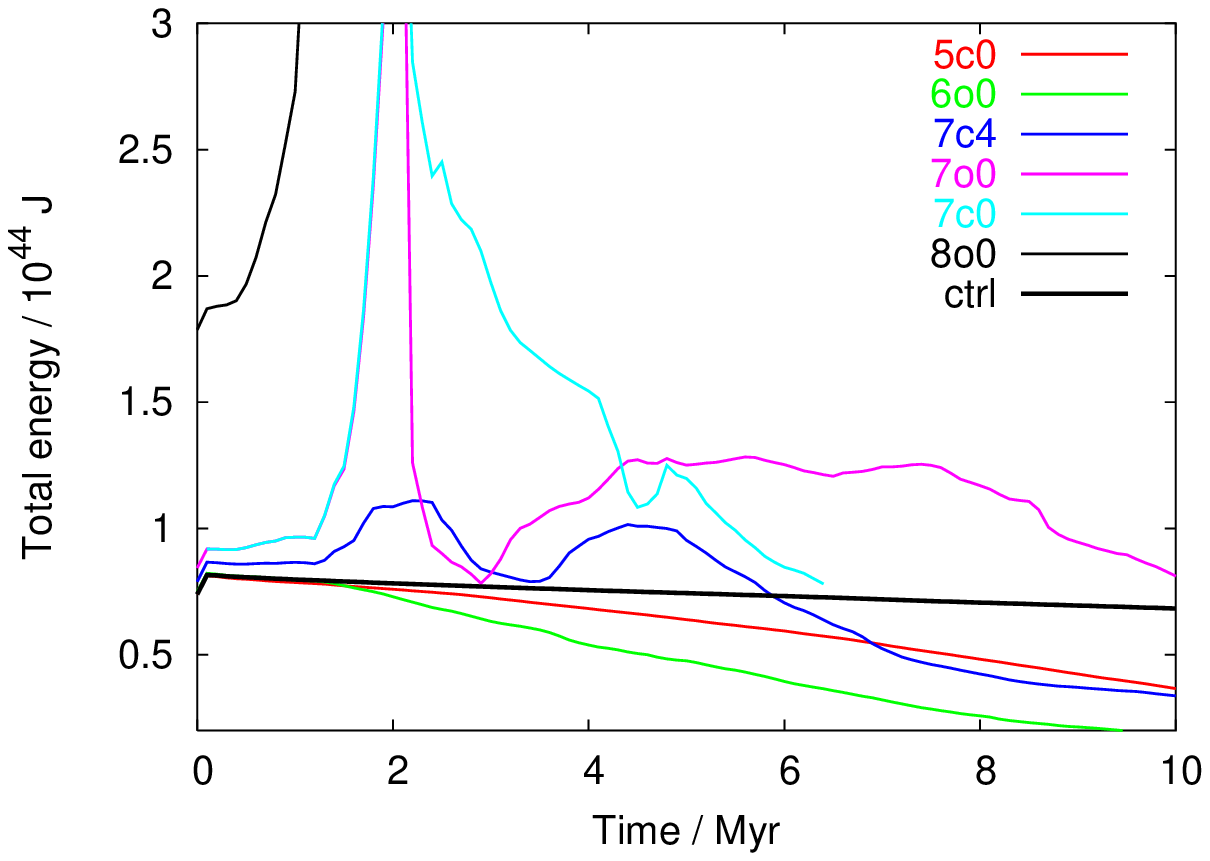}
\includegraphics[width=0.49\textwidth]{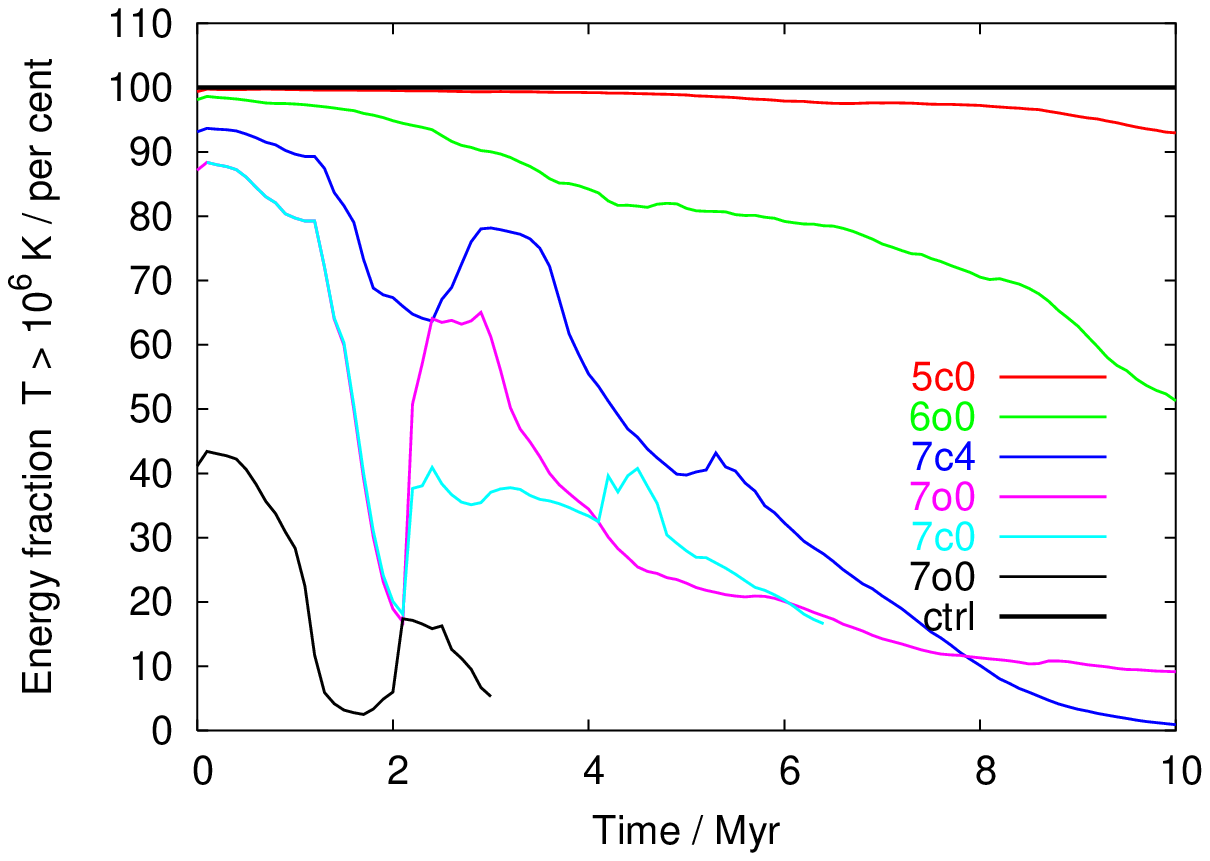}
\includegraphics[width=0.49\textwidth]{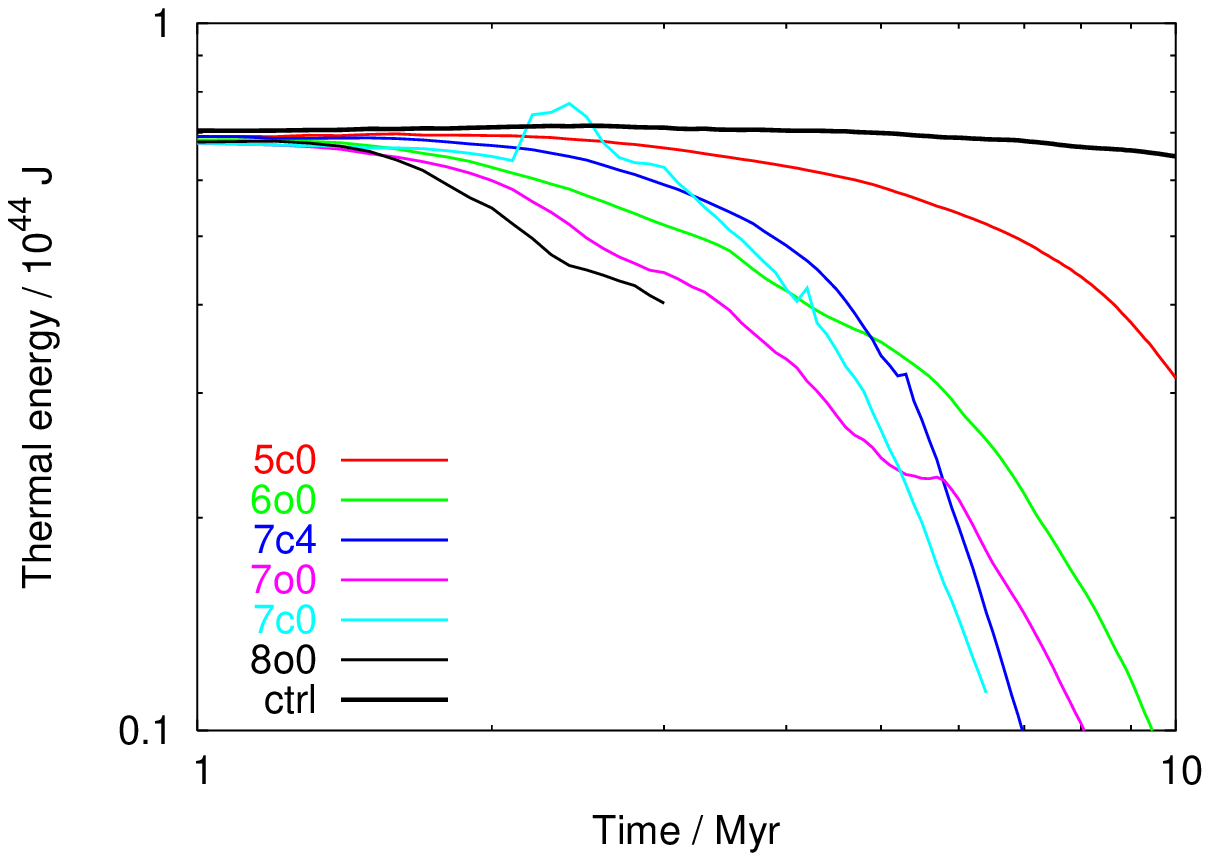}
\includegraphics[width=0.49\textwidth]{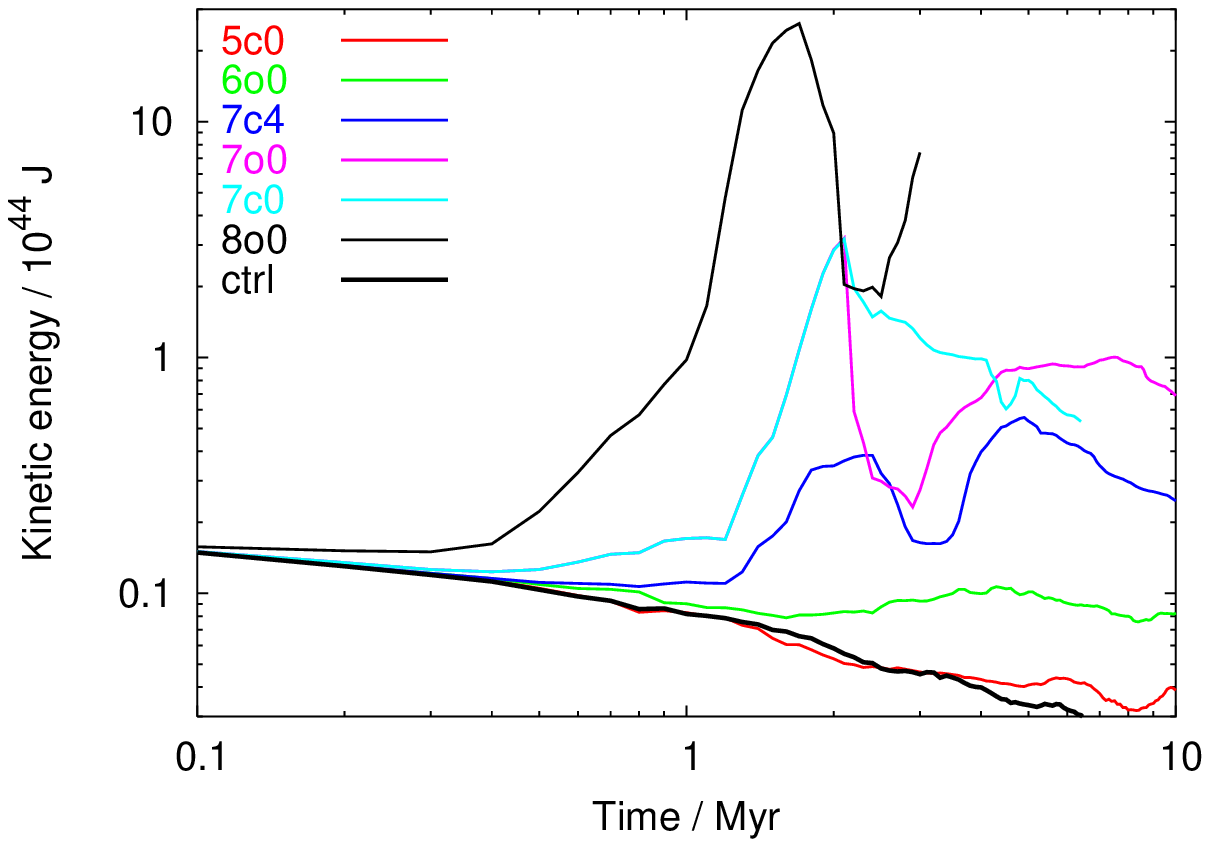}
\caption{Energy evolution over time for all runs. Top left: 
total (i.e. thermal, kinetic and gravitational) energy;
top right: fraction of total energy in gas hotter than $10^6$~K; 
bottom left: thermal energy;
bottom right: kinetic energy.
All runs besides ctrl, 5c0 and 6o0 show problems with energy conservation (see text), which can be
traced to the kinetic energy.}
\label{energy}
\end{figure*}
Energy conservation is not guaranteed formally by the numerical method employed by NIRVANA.
However, for practical considerations, it is usually conserved to reasonable accuracy, 
given sufficient resolution. This is the case for at least three of our runs 
(ctrl, 5c0, and 6o0, see Section~\ref{num.is} below for further discussion) --
we therefore first focus on the energy evolution of these three simulations. 
Figure~\ref{energy} (top left) shows the time evolution 
of the total energy for all runs, including the control run. Since there are no energy sources 
in the simulated volume, the total energy should monotonically decrease, at least in the 
closed box cases, due to the radiation term.
In the control run without clouds the system loses energy at a nearly constant rate
of $4\times 10^{28}$~W. For the simulation with the lightest clouds, the rate increases
to $18\times 10^{28}$~W, and for run 6o0, the total cooling increases further to   
$22\times 10^{28}$~W. As might have been expected, the presence of the cool clouds
increases the cooling rate. This is no longer evident for the simulations with higher 
mass loading. While at later times they turn to monotonic energy decrease at an even greater rate,
they are obviously dominated by numerical effects for some and in the case 
of run 8o0 even much of the simulation.
These numerical errors can be traced back to the kinetic energy. While the thermal energy
(Figure~\ref{energy}, bottom left) shows essentially a smooth energy decline, 
with the only significant dependence being on mass in the expected way, 
the kinetic energy displays strong and artificial wiggles, which is discussed below.
The extraction of energy from the high temperature system can be seen in yet another way.
The fraction of energy in gas hotter than $10^6$~K is shown in Figure~\ref{energy} (top right).
The system not only loses energy faster with increasing mass loading, but additionally 
energy is increasingly found in the low temperature gas.

\subsection{Cold mass dropout}
\begin{figure*}
\centering
\includegraphics[width=0.49\textwidth]{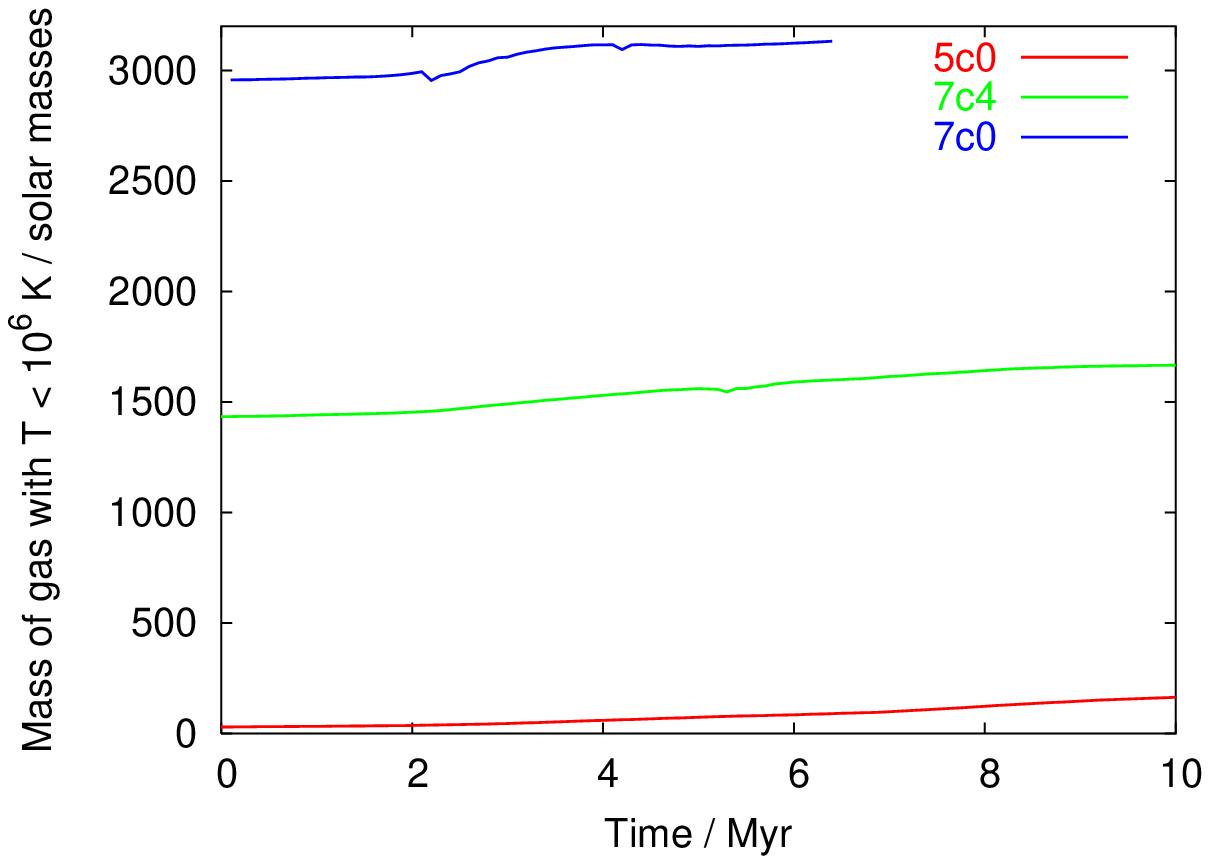}
\includegraphics[width=0.49\textwidth]{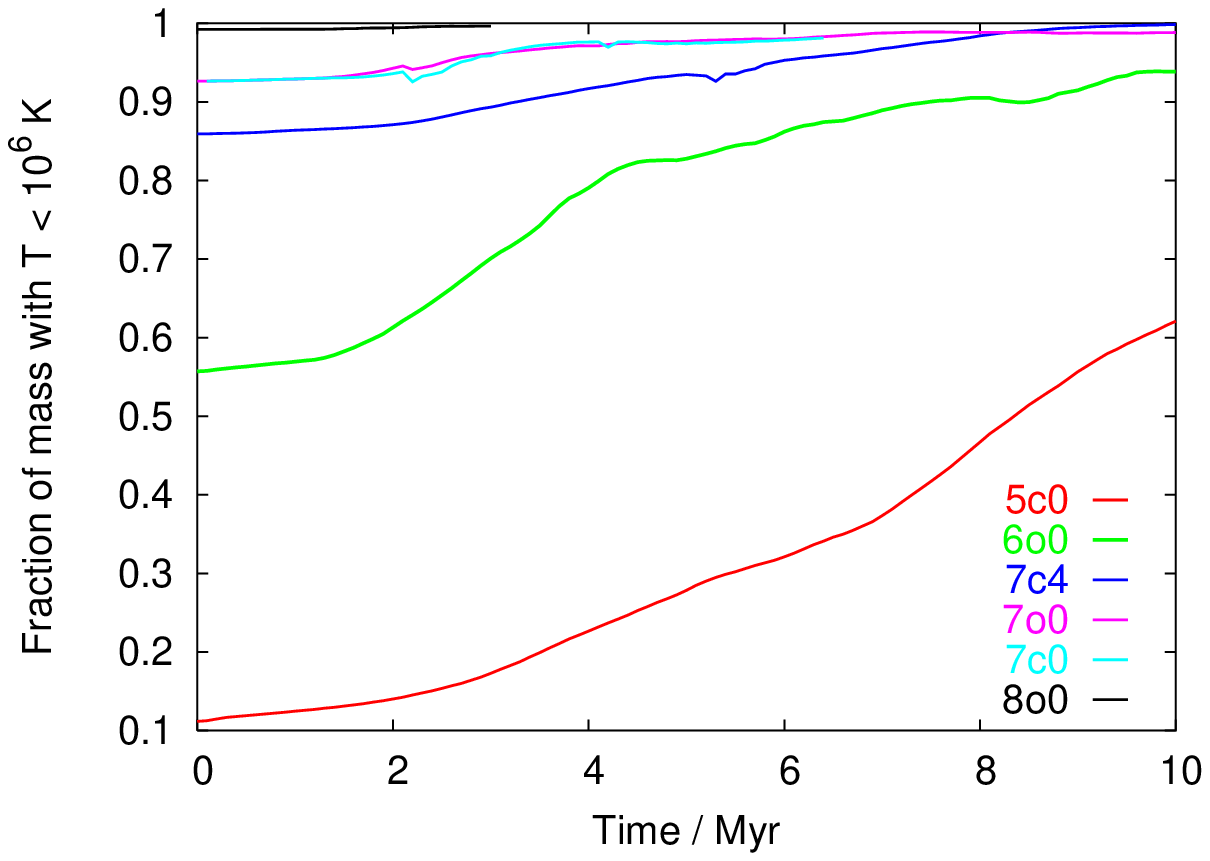}
\caption{Time evolution of the cold gas phase ($T< 10^6$~K) for cloud containing simulations. 
	Left: total amount of cold gas for the simulations with closed boundary.
	Right: Evolution of the cold gas mass fraction for all simulations.
	In the control run, no cold gas was detected over 10~Myr.}
\label{coldmass}
\end{figure*}
The increased system cooling rate due to the cold mass content leads to amplified mass 
dropout -- this is illustrated in Figure~\ref{coldmass}. As discussed in Section~\ref{ffm}, the 
total mass is conserved in the closed boundary simulations. The cold mass is therefore increased
by dropout from the warm phase, only. In the control run, there is no cold mass present 
initially, and none drops out during the simulation time.
Runs 5c0, 7c4, and 7c0 start with a cold 
mass of 30, 1434, and 2957~$M_\mathrm{\sun}$, respectively: they gain cold mass at rates of 
13.4, 23.3, and  27.3~$M_\mathrm{\sun}\,$Myr$^{-1}$, on average. 
The higher mass loaded simulations are limited 
by the available mass. At the beginning of the simulation, their cold gas mass is already
$\approx 90$ per cent of the total mass. They gain almost all of the rest during the simulation time.
The cold mass evolution for run 5c0 can be well fitted by an exponential growth with 
e-folding time of 6~Myr. This corresponds to a simple interpretation in which the rate of mass dropout
is proportional to the mass of cold gas. We note that runs 
7c0 and 7o0, which differ only in their boundary condition, have very similar cold gas mass fraction
histories, suggesting that the boundary conditions are unimportant for the 
fractional mass dropout. This is confirmed by the similar shape of all the curves in 
Figure~\ref{coldmass} (right). The simulations with small initial cold gas mass fraction
show increasing growth up to about 80 per cent. Further growth proceeds in a linear fashion.

\subsection{Numerical issues}\label{num.is}
As indicated above in Section~\ref{energyloss}, the higher mass loaded simulations show increasing
problems with energy conservation, in particular kinetic energy. Careful 
inspection of the simulations showed that the velocity errors depend strongly on the density
contrast between cells. In the high mass loaded simulations, as soon as the very light gas
in the lower half of the simulations comes into contact with the high density clouds, high 
velocity errors result. This is most evident in the total and kinetic energy plots 
(Figure~\ref{energy}). The higher the initial density, the higher the numerical error.
Simulations 5c0 and 6o0 show no indication of the problem. Run 7c4 shows moderate 
errors. Runs 7o0 and 7c0 display moderate errors with a singular spurious event around 2~Myr.
The fact that 7c4 shows only moderate errors at this time determines the critical density 
contrast to be about $10^7$. Run 8o0 is dominated 
strongly by numerical errors for much of the simulation time.  
We have checked that these errors decrease, as expected, if the numerical 
resolution is increased. Some aspects of the simulations are affected worse than others.
Regarding most of the results discussed so far, a noticeable similarity and continuation
of the results argues for the kinetic energy errors not to have a too 
large effect. 
In particular, the power spectrum is completely unaffected.
Comparison of the phase diagrams for runs 7o0 and 7c0 shows that once the gas 
which was 
accelerated in the singular spurious event has left the grid (7o0), the diagram 
soon reverts to normal, 
in contrast to the closed boundary simulation (7c0).
The high mass load simulations give the best signal to noise in the phase diagrams.
However, because of the energy conservation issue, we include run 8o0 only to demonstrate 
the limitation of the method most clearly, and base our conclusions exclusively 
on the very best results.

A further check we have performed was to repeat the simulation with the FLASH code
\citep{Fryxea00,Caldea02}, which guarantees energy conservation. However, in this case 
another problem appeared in low resolution runs. Here, the clouds were soon surrounded by 
a low pressure shell of a width equal to the resolution limit, which seriously damped the 
propagation of shocks into the clouds. 
The clouds essentially kept their shape for long times, even if accelerated. This problem 
could be solved by going to very high resolution, which corresponds for regions
of smooth flow roughly to 
a four times better resolution compared to the simulations presented above, 
due to the higher order scheme of FLASH.
\begin{figure}
\centering
\includegraphics[width=0.49\textwidth]{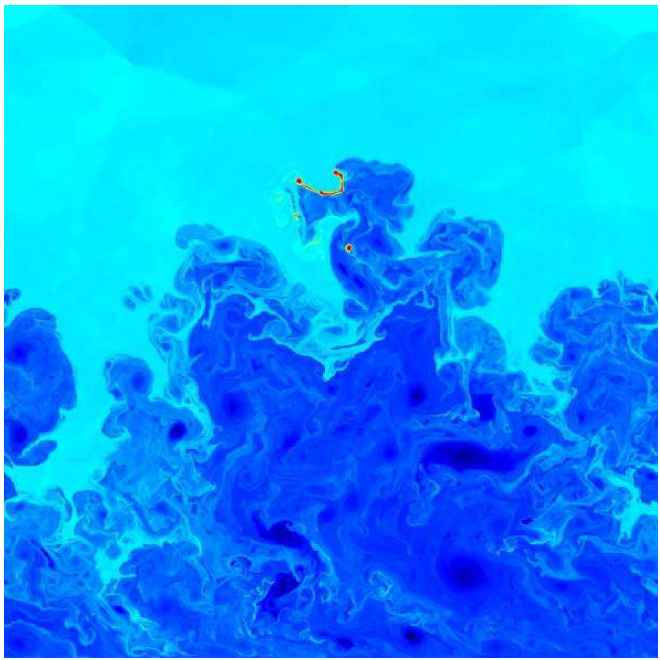}
\includegraphics[width=0.49\textwidth]{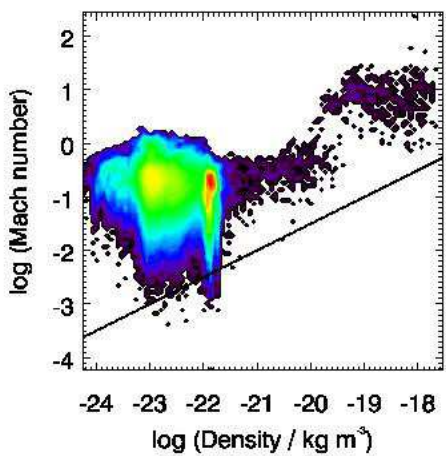}
\includegraphics[width=0.49\textwidth]{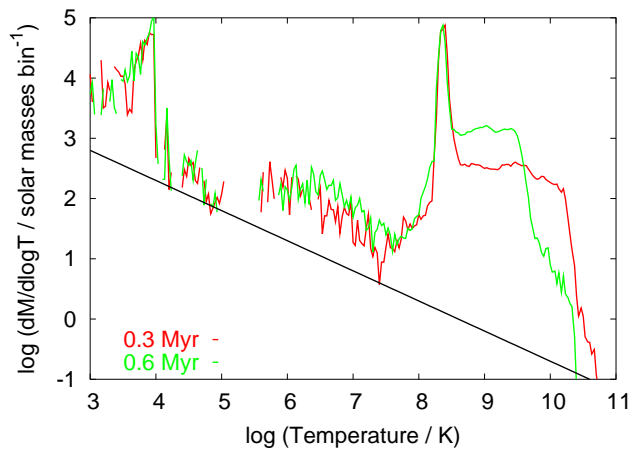}
\caption{Selection of results for the validation run with FLASH. 
	Top: logarithmic density distribution at 1~Myr, 
	middle: density Mach number histogram at 1~Myr, 
	bottom: gas mass over temperature, all gas, bottom right:
	gas mass over temperature between 5,000~K and 30,000~K, staggered for 13
	timesteps from 0.1~Myr to 1.3~Myr.}
\label{flash}
\end{figure}
We also adjusted the setup to have a single more massive cloud and allowed for higher background density
(top: $10^5 m_\mathrm{p}$~m$^{-3}$, bottom: $10^{2} m_\mathrm{p}$~m$^{-3}$). With such a setup we obtain very similar results
to the NIRVANA simulations (see Figure~\ref{flash}). The cloud soon collapses to form a filament, 
dispersing into little cloudlets afterwards. The Mach-number density relation is consistent 
with the NIRVANA results, and the gas mass versus temperature distribution shows the same general 
behaviour and structure.  

\subsection{Velocity distribution functions}
\begin{figure*}
\centering
\includegraphics[width=0.49\textwidth]{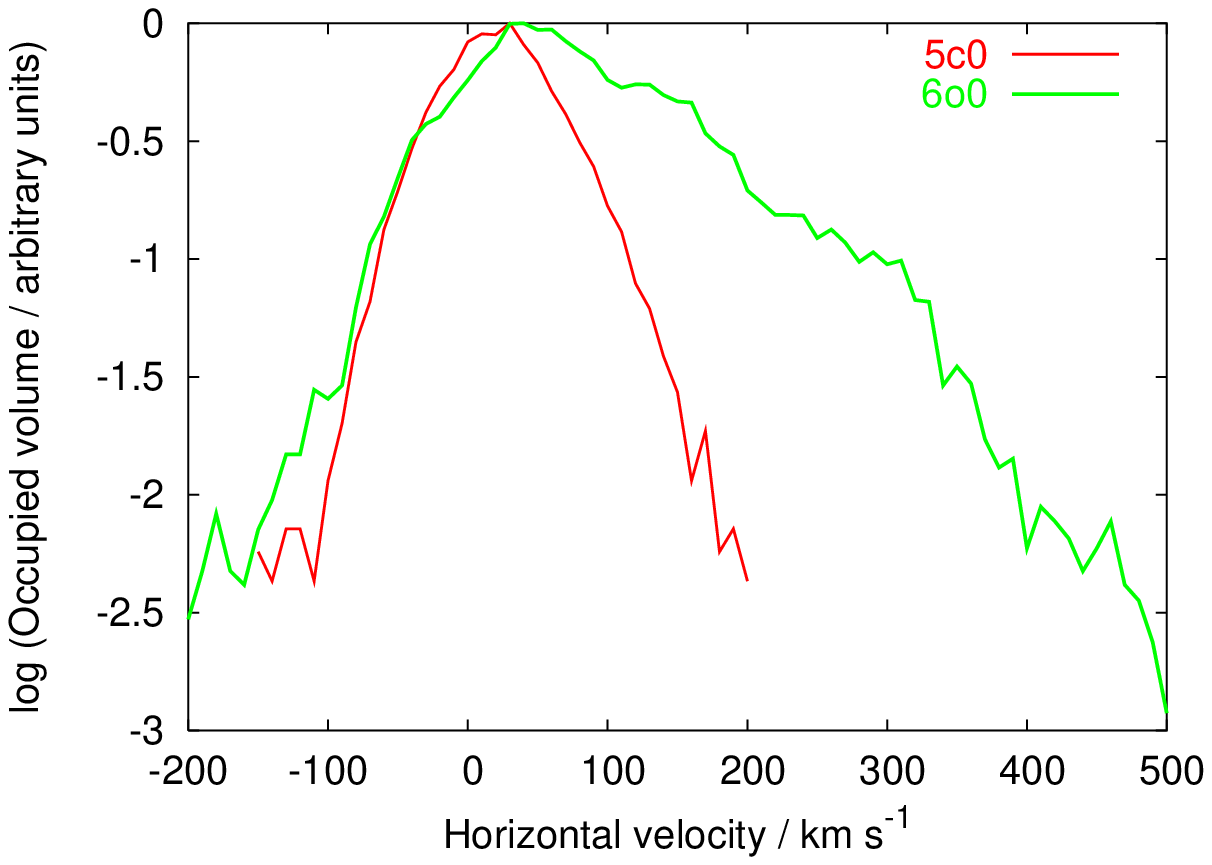}
\includegraphics[width=0.49\textwidth]{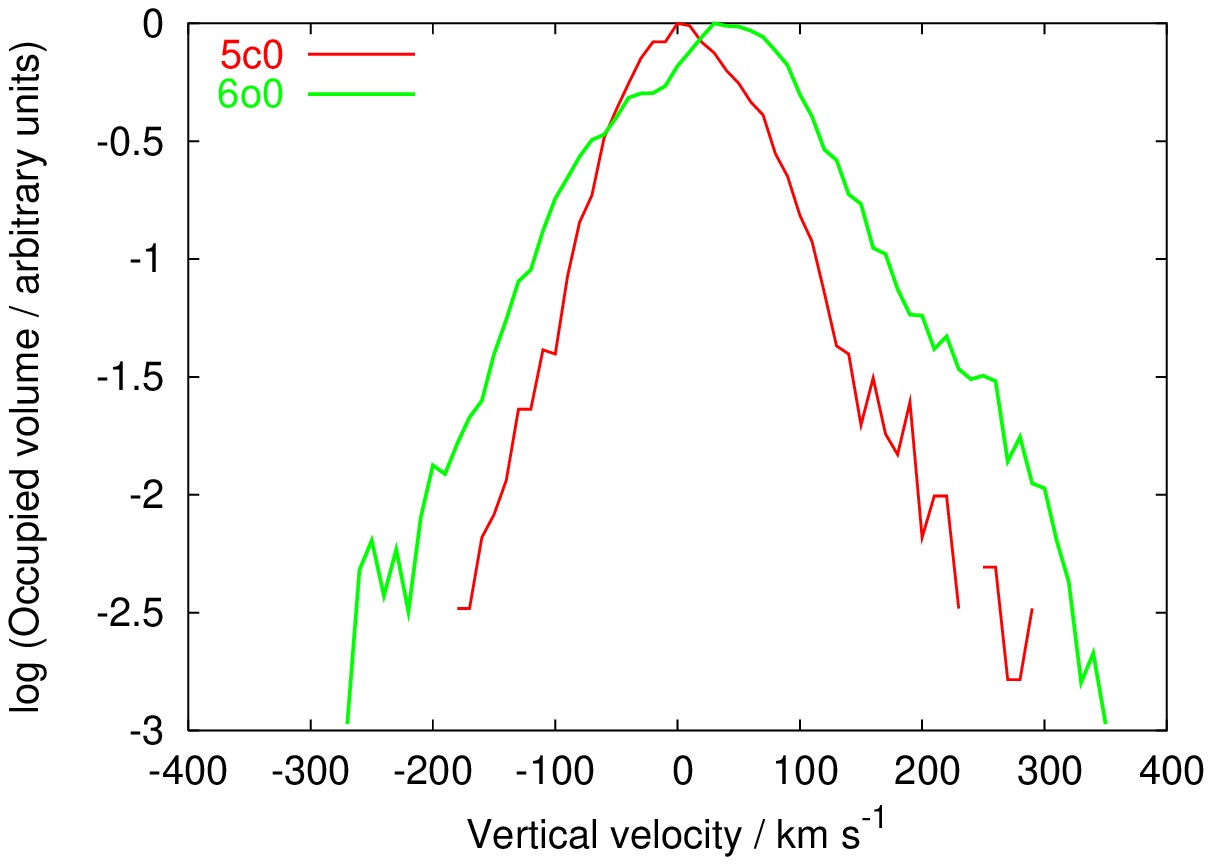}
\caption{Velocity distribution functions for gas at temperature 14,000~$\pm$~2,000~K 
	for runs 5c0 and 6o0.
	Left: horizontal velocity. Right: vertical velocity.
	For better signal to noise, output between 1 and 6~Myr (0.1~Myr steps) has been 
	staggered. The shift towards positive 
	horizontal velocities, more pronounced for the higher mass load case 
	is the result of the accelaration by the low density gas.}
\label{vpdf}
\end{figure*}

The velocity PDFs for the two most reliable runs (5c0 and 6o0) are shown in Figure~\ref{vpdf}.
As expected from general turbulence theory, they show
a Gaussian core with signs of acceleration due to drag and upwards driven shock waves
towards positive velocities.
These two most trustworthy simulations also show a dependence of the width on the mass 
loading. This confirms the earlier findings that in the simulations with higher mass load,
the fraction of kinetic energy in the cold phase rises faster.

\section{Discussion}\label{disc}

We present simulations of the Kelvin-Helmhotz instability with clouds
of differing density. The clouds are shocked, collapse into a filament, and then disperse 
into cloudlets and more filaments. There is no indication of the cold mass being heated 
beyond typical emission line gas temperatures.
On the contrary, over time more and more gas condenses into the cold phase.

Scale-free turbulence is establishes on the larger scales first.
The smallest 
scales remain anisotropic up to the end of the simulation

In general, the rms Mach number increases during the simulation. This is due to the 
cold gas aquiring high velocities. In particular, for gas with longer cooling time,
the Mach number is independent of the density, whereas for strongly cooling gas, 
the Mach number is
proportional to the square root of the density. The latter dependency is identical 
to the one found by \citet{KN04}.
 
The density probability distribution function shows a power law behaviour between 
the peaks imposed by the initial distribution, $P_\rho(\rho) \propto \rho^{-1/2}$. 
A similar behaviour is seen in the 
gas mass versus temperature plots. Generally, the distributions follow $M(T)\propto T^{-1/2}$,
with the region between $10^4$K and $10^6$K evolving gradually over a few million years
to follow the same scaling as at high temperature.
But what is the cause of this behaviour? Because of the roughly constant pressure,
the entropy distribution is also a power law. This rules out adiabatic processes.
We are left with shock heating, cooling and mixing. Cooling and shocks contribute 
little for the gas with $T>10^6$~K. 
So, mixing seems to be the best 
explanation for the power law slope of the distribution functions of density and 
temperature. Mixing is inevitable in multi-phase turbulence simulations. 
In reality, mixing is likely 
to occur at a lower level, consequently the enhancements over the mixing distributions 
should be more pronounced. There is a worry that the peak around 14,000~K described 
in the following might be entirely due to the mixing.
Against this possibility argues:
\begin{itemize}
 \item The peak builds up early, when the mixing in the cold gas is still at low level.
 \item The 14,000~K peak joins the lower temperature gas above the mixing power law.
 \item The much less diffusive Flash simulation, produces a very similar peak. 
\end{itemize}
We conclude that the 14,000~K peak is probably unaffected by the mixing.

This peak is in fact likely to be due to shock heating combined with the effects of the 
steep cooling curve. The peak appears in all simulations at most times.
Since there is a drop in the cooling function towards lower temperature and there is a drop in gas 
mass towards higher temperature, the gas close to this temperature may contribute most of the optical 
emission. Its filling factor depends on the initial mass loading, and corresponds to $10^{-5}$
to $10^{-3}$, for a realistic 3D generalisation. The velocity of this gas is
distributed like a Gaussian in run 5o0. Run 6c0 has a pronounced deviation
on the positive side due to acceleration, and appears unrelaxed. 
The typical width of observed 
emission lines may be understood 
by the Mach-number density relation. Gas with low cooling time was shown to have $M \propto \sqrt{\rho}$,
i.e. the velocity is constant. This is valid for gas of a temperature up to 
$\approx10^6$~K. For hot gas, the Mach number is constant, i.e the velocity drops with $\rho^{-1/2}$.
The hot radio plasma, has velocities of typically $10^5$~km$\,$s$^{-1}$ when entering the cocoon, which will 
become increasingly turbulent. If the entrained ambient gas is roughly 10,000 times denser, it would be
accelerated to typical velocities of 1000~km$\,$s$^{-1}$. If in $z\approx 1$ radio galaxies
that gas has a temperature of $\ga 10^7$~K as for nearby sources, then gas with $T<10^6$~K
would have a typical velocity width of a few 100~km$\,$s$^{-1}$. However, the typical observed velocity
width is $\approx 1000$~km$\,$s$^{-1}$. Hence, either the density ratio in these sources is 
typically less extreme than assumed here, 
or the typical temperature of the ambient gas is below $10^7$~K  
(for both, a factor of $\approx 10$ would suffice).

Excess cold mass 
is located in a broad peak around $10^3$K, which would contain substantial 
amounts of molecular hydrogen. Most of the gas mass would be at this temperature,
unless there is heating by photoionisation. 

Higher mass loaded simulations extract the energy much more efficiently from the warm and hot 
phase. Therefore, the velocity widths as well as the energy extraction increase with mass loading.
The overall cooling rate of the system was increased by about a factor of five for the low
mass loaded simulations. For the high mass loaded simulations it was shown to increase 
further, whereas the quality of the simulations gets worse.

The simulations demonstrate the loss of numerical accuracy with increasing density 
contrasts. Three of the simulations conserve energy to a very good accuracy;
further three have moderate problems, whereas the highest mass loading simulation 
has significant errors with respect to the energy conservation. This is a clear trend.
We conclude that the highest density contrast in a simulation is the limiting factor
for numerical reliability.
Despite these problems at extreme density ratios we believe the main trends and results are 
very reliable -- we 
have checked this by comparing results to a high-resolution simulation performed using the 
FLASH
code which is forced to be energy conserving.

The cold gas clouds effectively act as condensation nuclei. Strong growth of the cold gas mass
is observed in all simulations which is in contrast to the control run in which no gas cools 
below $10^6$K for the whole simulation time. The lowest gas mass run, which is furthest from 
saturation, gains a factor of more than five in cold gas mass within 10~Myr. In a real radio 
cocoon the supply of warm gas would be ample, and exponential growth would result.
Hence, a cold gas mass seed of order 100~$M_\mathrm{\sun}$ would be sufficient to condense 
the observed $\approx 10^9 M_\mathrm{\sun}$ in the radio galaxies at $z \approx 1$ mentioned 
in the introduction, provided the conditions are similar to those considered here 
and if the typical source age is of order 100~Myr. It is worth noting
that $\approx 10^9 M_\mathrm{\sun}$ is also the amount of X-ray gas entrained into the 
radio cocoon as infered from simulation \citep{Krause2005a}. 
A consistent picture emerges, if most of the entrained X-ray gas 
cools to the cold phase via the help of a small cold-gas seed. As this gas cools, most 
of the radiated energy is not in the 
X-ray band, but in the optical. Since in the real radio cocoon, unlike in the simulation here,
the most energetic part of the system would be the radio emitting hot phase, a tight 
correlation between radio and optical emission would be expected, which is indeed observed 
\citep[e.g.][]{MC93}.
Such a mechanism also of course explains the increased visibility of optically emitting 
regions after the passage of the radio cocoon.

Due to the large temperature ratios, heat conduction is a particular concern. 
However, it may still be negligible here, as we argue in the following.
First, the evaporation lengthscale \citep{BF89} is at least a factor of thousand below
our resolution limit. Second, using Spitzer conductivity, we can estimate the energy transfer
per volume due to heat conduction:
\begin{displaymath}
\dot{e}= \nabla \cdot [\sigma T^{2.5} \nabla T] \approx \frac{\sigma T^{3.5}}{dx^2} 
	\approx 10^{-23} \mathrm{W}\,\mathrm{m}^{-3} \left(\frac{T}{10^6\,\mathrm{K}}\right)^{3.5}.
\end{displaymath}
This leads to a heat conduction timescale of:
\begin{displaymath}
\tau_\mathrm{hc}=\frac{e}{\dot{e}}\approx 
	\frac{10^{-13}\mathrm{J}\,\mathrm{m}^{-3}}
	     {10^{-23} \mathrm{W}\,\mathrm{m}^{-3} (T/10^6\,\mathrm{K})^{3.5}}
	= 10^{10}\,\mathrm{s} \left(\frac{T}{10^6\,\mathrm{K}}\right)^{-3.5}.
\end{displaymath}
Since our simulations represent a dynamical equilibrium, we should compare this to the 
cooling timescale in order to find out to what extent heat conduction 
could contribute. The cooling time for the hot, warm and cold gas is of order $10^{20}$~s,
$10^{15}$~s and $10^{11}$~s, respectively. So, heat conduction from the hot phase would 
clearly dominate everything else. However, in reality the presence of magnetic fields
strongly suppresses heat conduction in and from the hot phase. If this was not the case,
radio cocoons could not exist, since the internal energy would have been lost on the light
crossing timescale. Among the cool gas, the heat conduction timescale is too long to be
of concern. Therefore, we are left with the heat flow from the warm to the cold phase.
Here, the heat conduction is about 10 times faster than the radiative cooling.
For the simulation parameters, saturation sets in just above $10^6$~K
\citep{CMcK77}, and does therefore not influence the conclusion.
However, also in the warm gas, the heat conduction is severly suppressed due to the 
presence of magnetic fields, at least for the current cosmological age.
Measurements from X-ray data suggest a suppression by one to three orders of magnitude
\citep{EF00,Markea03,Nath03}.
The magnetic field in the warm gas around high redshift radio sources is unknown.
However, if its effect on heat conduction is similar than in nearby cases, it will 
probably be a minor contribution to the total heat transfer.
If the heat conduction would be less strongly suppressed, the processes described in this 
paper would be further amplified - more heat would be radiated by the cold phase
and more warm gas would condensate on the cool gas.

The simulations presented here are 2D, only. This is an important first step, and it is 
very likely 
that the mechanisms we discuss would also be present in 3D -- the Mach-number
density relation is identical to published 3D results. To get more accurate estimates 
for example for
cooling and mass dropout amplification or filling factors, 3D simulations are required.
This is likely to be challenging, given the importance of high resolution,
but will no doubt be possible in the near future.

\section{Conclusions}
We presented 2D turbulence simulations initiated and driven only by the 
Kelvin-Helmholtz instability. We add dense, cooling clouds to the simulations, so that 
the interaction of the three phases present in the cocoons of high redshift radio 
galaxies may be examined. We find a turbulent cascade with energy spectrum proportional
to $k^{-5/3}$ at larger scales, as in Kolmogorov turbulence. 
Mach number and density show a bimodal correlation with a break at a 
characteristic density, where the cooling time is about the simulation time.
In the high density part, the Mach number goes as $\rho^{1/2}$, which is the same
as found by independent 3D simulations by \citet{KN02}.
We use the relation to infer from the observed kinematics that either the 
density ratio in the high redshift sources is different from their low redshift counterparts
or the temperature in the environment is lower.
The temperature distribution function is dominated by a power law which we ascribe mainly
to mixing. We find a strong peak in the distribution function at 14,000~K. This coincides
with a strong rise in the cooling function, and corresponds to an equilibrium
between shock heating and adiabatic compression on the one side and radiative cooling
on the other side. We suggest that such gas is responsible for the optical emission
in high redshifted radio galaxies, whenever shock ionisation dominates over photoionisation.
We find a filling factor of $10^{-5}$ to $10^{-3}$ for the gas in this peak.
The velocity width for this gas increases with mass load and is about 100~km/s for
the simulations we consider to be reliable in this respect.
By comparison of simulations with different loads of cold mass, we establish that 
the cooling time for the hotter phases is considerably reduced by the presence of the
cold gas. We find an exponential growth of the cold mass with time due to cooling of 
warmer gas. The growth rate is sufficient to explain the origin of the optical gas 
associated with the cocoons of radio galaxies, as being build up by this 
turbulence enhanced cooling process. Heat conduction is not able to evaporate the 
cool clouds we inject. However, in cases where the suppression below the Spitzer value
is less extreme than deduced from observations in nearby sources, it might contribute
to the heat transfer significantly, and would presumably amplify the process we describe.

\section*{Acknowledgments}
We thank the referee for thorough reading of the manuscript and very helpful suggestions.
MK acknowledges a fellowship from the Deutsche Forschungsgemeinschaft (KR 2857/1-1)
and the hospitality of the Cavendish Laboratory, where this work has been carried out.
The software used in this work was in part developed by the 
DOE-supported ASC / Alliance Center for Astrophysical Thermonuclear Flashes 
at the University of Chicago.

\appendix

\bsp
\bibliographystyle{apj}
\bibliography{/home/krause/texinput/references}

\label{lastpage}

\end{document}

%% file: def.tex
\bibpunct{(}{)}{;}{a}{}{,}

\newcommand{\ignore}[1]{}
\newcommand{\eq}[1]{\begin{equation} #1 \end{equation}}

\newcommand{\der}[2]{\frac{\partial #1}{\partial #2}}

\newcommand{\dt}[1]{\der{#1}{t}}

\newcommand{\dm}[1]{\begin{displaymath} #1 \end{displaymath}}